\documentclass[aps,prd,showpacs,superscriptaddress,preprint,raggedfooter,raggedbottom]{revtex4-1}   

\usepackage{amssymb}
\usepackage{amsmath}
\usepackage{amssymb}
\usepackage{graphicx}
\usepackage{subfigure}
\usepackage{color}
\usepackage{mathrsfs} 

\usepackage{soul}

\usepackage{hyperref}

\begin{document}

\title{Long-range interactions of kinks}
\author{Ivan C.\ Christov}
\affiliation{School of Mechanical Engineering, Purdue University, West Lafayette, Indiana 47907, USA}
\author{Robert J.\ Decker}
\affiliation{Mathematics Department, University of Hartford, 200 Bloomfield Ave., West Hartford, CT 06117, USA}
\author{A.\ Demirkaya}
\affiliation{Mathematics Department, University of Hartford, 200 Bloomfield Ave., West Hartford, CT 06117, USA}
\author{Vakhid~A.~Gani}
\affiliation{Department of Mathematics, National Research Nuclear University MEPhI (Moscow Engineering Physics Institute), 115409 Moscow, Russia}
\affiliation{Theory Department, National Research Center Kurchatov Institute, Institute for Theoretical and Experimental Physics, 117218 Moscow, Russia}
\author{P.~G.\ Kevrekidis}
\affiliation{Department of Mathematics and Statistics, University of Massachusetts,
Amherst, MA 01003-4515, USA}
\author{R.~V.\ Radomskiy}
\affiliation{P.~N.\ Lebedev Physical Institute of the Russian Academy of Sciences, Moscow 119991, Russia}

\begin{abstract}
We present a computational analysis of the long-range interactions of solitary waves in higher-order field theories. Our vehicle of choice is the $\varphi^8$ field theory, although we explore similar issues in example $\varphi^{10}$ and $\varphi^{12}$  models. In particular, we discuss the fundamental differences between the latter higher-order models and the standard $\varphi^4$ model. Upon establishing the power-law asymptotics of the model's solutions' approach towards one of the steady states, we make the case that such asymptotics require particular care in setting up multi-soliton initial conditions. A naive implementation of additive or multiplicative ans{\"a}tze gives rise to highly pronounced radiation effects and eventually leads to the illusion of a repulsive interaction between a kink and an antikink in such higher-order field theories. We propose and compare several methods for how to ``distill'' the initial data into suitable ans{\"a}tze, and we show how these approaches capture the attractive nature of interactions between the topological solitons in the presence of power-law tails (long-range interactions). This development paves the way for a systematic examination of solitary wave interactions in higher-order field theories and raises some intriguing questions regarding potential experimental observations of such interactions. As an Appendix, we present an analysis of kink-antikink interactions in the example models via the method of collective coordinates. 
\end{abstract}

\maketitle


\section{Introduction}

Field-theoretic models with polynomial potentials are of great interest in many areas of modern theoretical physics: from cosmology \cite{vilenkin01,manton} to condensed matter \cite{kiselev,Bishop80}. For example, the scalar $\varphi^4$ model with two minima of the potential is widely used to model spontaneous symmetry breaking. Besides that, the quartic potential arises in the phenomenological Ginzburg--Landau model of superconductivity \cite{GinzburgLandau,Tinkham}; see also Refs.~\cite{Bishop80,csw} for an overview of different areas of application. In this setting, the dynamics of the complex scalar field of Cooper pairs is described by a polynomial self-interaction of the fourth degree. Models with polynomial potentials of higher degrees are commonly used, e.g.\ to model consecutive phase transitions \cite{khare}, which arise in material science \cite{gl}, as recently summarized in \cite{HOFT_chapter}. It has also been shown that scalar field theories can describe distinct quantum mechanical problems (including supersymmetric ones) \cite{bazeia17}, leading to new applications of field theories of the type $\varphi^{2n}$. Another possibility is to consider scalar $\varphi^{2n}$ field theories as Lane--Emden truncations of a periodic potential, which then leads to applications of these theories as toy models for dark matter halos \cite{valle2016}. Field-theoretic models with polynomial potentials that can exhibit topological solutions (kinks)  are also important in cosmological applications of the Higgs field \cite{poltis1,poltis2}. { Beyond field theoretic models with polynomial potentials, finite-gap potentials of Lam\'e type also lead to scalar field theories with exotic kink solutions, now relevant in the context of sypersymmetric quantum mechanics  \cite{Plyushchay.PRD.2014,Plyushchay.PRD.2015} and extended to $\mathcal{PT}$-symmetric situations \cite{Plyushchay.JHEP.2017}.}

Although the above-mentioned models { with polynomial potentials} are non-integrable, studying their properties in (1+1)-dimensional space-time is of common interest because, in this setting, a variety of analytical (and numerical) methods can be straightforwardly deployed to fully understand the dynamics of coherent structures. Moreover, (1+1)-dimensional solutions may be relevant to more realistic situations in higher dimensions; for example, the equations for certain five-dimensional brane-world phenomenologies can be reduced to differential equations similar to those of (1+1)-dimensional field theories \cite{DeWolfe}. Such models with polynomial potentials of even degree allow kinks --- topological solutions that interpolate between neighboring minima of the potential, i.e.\ vacua of the model \cite{vach06}. Properties of kinks of the $\varphi^4$ and $\varphi^6$ models are well-studied, yielding many important results \cite{Bishop80,csw,anninos,goodman2,dorey,Campbell1,goodman,GaKuLi,weigel01,weigel02,simas01,MGSDJ,DDKCS,Romanczukiewicz.PLB}. At the same time, polynomial potentials of higher degrees have not been studied systematically. Nevertheless, the work has been started \cite{lohe,bazeia06-11-13,Gomes.PRD.2012,khare,GaLeLi,GaLeLiconf,gani17,Belendryasova.conf.2017,Belendryasova.arXiv.08.2017,HOFT_chapter}. In particular, exact (but implicit) solutions for the kink shapes of various $\varphi^8$, $\varphi^{10}$ and $\varphi^{12}$ field theories have been obtained \cite{khare}, the excitation spectra of the $\varphi^8$ kinks have been studied, resonance phenomena in the kink-antikink scattering have been found, and their relation to the kinks' vibrational excitations has been discussed \cite{GaLeLi,GaLeLiconf,Belendryasova.conf.2017,Belendryasova.arXiv.08.2017}. {As described below, we use the purely theoretical foundation established in \cite{khare} to develop a novel understanding of long-range kink interactions, re-assessing in significant detail the interaction picture first put forth in \cite{Belendryasova.conf.2017,Belendryasova.arXiv.08.2017}. However, issues of vibrational modes in higher-order field theory \cite{GaLeLi,GaLeLiconf} remain beyond the distinct scope of the present work.}

Until recently, the dynamics of kink-(anti)kink interactions had been studied only for kinks with exponential tail asymptotics. At the same time, it is easy to show that, for certain potentials of sixth or higher degree, there exist kinks that exhibit power-law asymptotics of either or both tails connecting two distinct equilibria. Although conditions for algebraic soliton solutions of the nonlinear Schr\"odinger, Korteweg--de Vries and related integrable models have long been known \cite{Kaup77,Ablowitz77}, the case of kinks with algebraic tails in non-integrable $\varphi^{2n}$ field theories remain less explored in comparison.

In this paper, building upon the preliminary report of some of the present authors~\cite{gani17}, we systematize restrictions on the potential and obtain general formul\ae\ for the kinks' tail asymptotics (see also Refs.~\cite{khare,lohe,Gomes.PRD.2012,Bazeia18}), including the conditions for power-law tail asymptotics in a sextic potential. The existence of kinks with power-law tails is of particular interest because such tails lead to \emph{long-range interaction} between a kink and an antikink. Studying such long-range interactions at the effective ``particle'' level is a topic of significant current interest. Thus, we undertake and exploration of the interaction chiefly via direct numerical simulations of the relevant field equation. An Appendix complements our computations with analytical considerations based on the variational technique known as the \emph{collective coordinate approximation}, which is widely used in various field-theoretic problems, see, e.g.\ \cite{Sugiyama,csw,aek01,malomed,GaKuLi,weigel01,weigel02,baron01,berenstein,hata01,javidan,shnir,DDKCS}. We note, in passing, that other approaches have been used to interrogate long-range interactions, such as evaluating the field's potential energy at the center of mass of two superimposed (anti)kinks~\cite{Gonz1,Gonz2}. Also, the effect of minima of the potential and their relative depth (including inflection points in-between) on the kink asymptotics and their mutual forces was considered in~\cite{Gonz3}. An alternative method for identifying the interactions of solitary waves is the so-called Manton's method that has been widely used in solitonic equations~\cite{manton,manton_npb,perring62,rajaram77,kks04}.
{A very recent attempt to utilize this and other methods to infer a power-law dependence of the force on the kink-antikink separation, in the case of the $\varphi^8$ model, has just been posted { in~\cite{mantonnow,mantonnow2}.} The result in { \cite{mantonnow}} corroborates our previous observation in \cite{gani17} that, for an example eight-order potential with three degenerate minima and one-sided power-law tail asymptotics of the kink, a kink and an antikink attract each other with a force proportional to their separation to the $-4$ power. In this work, we provide numerical evidence for this type of power-law attractive force. However, a more conclusive theoretical investigation (including discussion of the pre-factor on this power law) is deferred to future work; see also the relevant discussion in Sec.~\ref{sec:conclusions}.}
  
Our presentation is structured as follows. In Sec.~\ref{sec:analytical}, we begin by providing the kink asymptotics in higher-order $\varphi^{2n}$ field theories (and specifically for our principal $\varphi^8$ example).  Subsequently, in Sec.~\ref{sec:num}, we delve into our numerical considerations, starting with how numerical experiments of kink-antikink collisions are set up, explaining the difficulties of such a setup for higher-order field theories, and proposing a corresponding methodology for handling such difficulties in the $\varphi^8$ model. The latter are complemented by parallel considerations of example $\varphi^{10}$ and $\varphi^{12}$ field theories. A discussion of the force of interactions between a kink and an antikink via power-law tails is presented in Sec.~\ref{sec:force}. We then summarize our work and propose some (among the many intriguing) questions for future work in Sec.~\ref{sec:conclusions}. In the Appendix, we explain how to perform a calculation of long-range interactions based on the method of collective coordinates.

\section{Analytical Considerations}
\label{sec:analytical}

\subsection{Power-law asymptotics of kinks}
\label{sec:Asymptotics}

Consider a real scalar field $\varphi(x,t)$ in $(1+1)$ dimensional space-time, with its dynamics determined by the Lagrangian density
\begin{equation}\label{eq:largang}
	\mathscr{L}=\frac{1}{2} \left( \frac{\partial\varphi}{\partial t} \right)^2-\frac{1}{2} \left( \frac{\partial\varphi}{\partial x} \right) ^2-V(\varphi),			
\end{equation}
where $V(\varphi)$ is a potential that defines the self-interaction of the field $\varphi$. The energy functional corresponding to the Lagrangian (\ref{eq:largang}) is
\begin{equation}\label{eq:energ}
	E[\varphi]=\int_{-\infty}^{+\infty}\left[\frac{1}{2} \left( \frac{\partial\varphi}{\partial t} \right)^2+\frac{1}{2} \left( \frac{\partial\varphi}{\partial x} \right) ^2+V(\varphi)\right]dx.
\end{equation}
Assume that the potential $V(\varphi)$ is a non-negative polynomial of even degree (denoted in short-hand as ``$\varphi^{2n}$'') having two or more minima $\bar{\varphi}_1$, $\bar{\varphi}_2$, $\hdots$, $\bar{\varphi}_n$ of equal depth  $V(\bar{\varphi}_1)=V(\bar{\varphi}_2)=\cdots=V(\bar{\varphi}_n)=0$. Consider two adjacent minima $\bar{\varphi}_i$ and $\bar{\varphi}_{i+1}$. Let $\varphi_\mathrm{K}^{}(x)$ be a kink (see, e.g.\ \cite[Ch.~5]{manton}) interpolating between these minima, i.e.,
\begin{equation}
\lim\limits_{x\to -\infty}\varphi_\mathrm{K}^{}(x)=\bar{\varphi}_i, \qquad \lim\limits_{x\to +\infty}\varphi_\mathrm{K}^{}(x)=\bar{\varphi}_{i+1}.
\end{equation}
According to conventional notation, we can say that this kink belongs to the topological sector $(\bar{\varphi}_i,\bar{\varphi}_{i+1})$, and we denote it by $\varphi_{(\bar{\varphi}_i,\bar{\varphi}_{i+1})}(x)$.

The Euler--Lagrange equation of motion, which is the condition for extremizing the action generated by the Lagrangian density \eqref{eq:largang}, is
\begin{equation}\label{eq:nkg}
	\frac{\partial^2\varphi}{\partial t^2} = \frac{\partial^2\varphi}{\partial x^2} - V'(\varphi).
\end{equation}
The kink shape function $\varphi=\varphi_\mathrm{K}^{}(x)$ is a \emph{time-independent} solution of Eq.~\eqref{eq:nkg}, and it can be shown that it satisfies a first-order ordinary differential equation:
\begin{equation}\label{eq:bps}
\frac{d\varphi}{dx}=\sqrt{2V(\varphi)}.
\end{equation}
A static field configuration that satisfies Eq.~\eqref{eq:bps} has the minimal energy among all the configurations in a given topological sector. The solutions that satisfy Eq.~\eqref{eq:bps} are called BPS-saturated configurations (BPS standing for Bogomolny--Prasad--Sommerfield)~\cite{bps}. Moving kinks can be obtained from solutions of Eq.~\eqref{eq:bps} via a boost transformation owing to the Lorentz invariance of the field theory. Using Eq.~\eqref{eq:bps}, the energy \eqref{eq:energ} can be rewritten as
\begin{equation}
E[\varphi]=\int_{-\infty}^{+\infty} \sqrt{2V(\varphi)}\cdot \sqrt{2V(\varphi)}\:dx.	
\end{equation}
Taking into account that $\sqrt{2V(\varphi)}\:dx=d\varphi$, the energy (rest mass) of a static BPS-saturated field configuration is
\begin{equation}\label{eq:stenerg_bps}	M=\int_{\bar{\varphi}_{i}}^{\bar{\varphi}_{i+1}} \sqrt{2V(\varphi)}\:d\varphi.
\end{equation}

Now, we formulate general conditions that must be satisfied in order for the model to have kinks with power-law tail asymptotics. We also give general formul\ae\ for such asymptotics. Let us turn our attention to the potential $V(\varphi)$. Let $\bar{\varphi}_i$ and $\bar{\varphi}_{i+1}$ be zeros of the function $V(\varphi)$ of orders $k_i$ and $k_{i+1}$, respectively. Notice that $k_i$ and $k_{i+1}$ must be even. We assume $\bar{\varphi}_i<\bar{\varphi}_{i+1}$, for definiteness. Then, the potential $V(\varphi)$ can be written as
\begin{equation}\label{eq:potzeros}
V(\varphi)=\left(\varphi-\bar{\varphi}_i\right)^{k_i}\left(\varphi-\bar{\varphi}_{i+1}\right)^{k_{i+1}}V_1(\varphi),
\end{equation}
where $V_1(\bar{\varphi}_i)>0$, $V_1(\bar{\varphi}_{i+1})>0$. Inserting Eq.~\eqref{eq:potzeros} into Eq.~\eqref{eq:bps} and recalling that $\bar{\varphi}_i<\varphi<\bar{\varphi}_{i+1}$, we obtain after integrating:
\begin{equation}\label{eq:int_bps}
\int dx=\int\frac{d\varphi}{\left(\varphi-\bar{\varphi}_i\right)^{k_i/2}\left(\bar{\varphi}_{i+1}-\varphi\right)^{k_{i+1}/2}\sqrt{2V_1(\varphi)}} \, .
\end{equation}
To find asymptotics of the kink $\varphi_{(\bar{\varphi}_i,\bar{\varphi}_{i+1})}(x)$ as $x\to -\infty$, we use the fact that $\varphi\to\bar{\varphi}_i$ in this limit. Then, there are slowly varying factors within the integrand in the right-hand side of Eq.~\eqref{eq:int_bps} at $\varphi\to\bar{\varphi}_i$. These factors can be (approximately) taken out from the integral. As a result, we obtain an asymptotic equality:
\begin{equation}\label{eq:int_x}
\int dx\approx\frac{1}{\left(\bar{\varphi}_{i+1}-\bar{\varphi}_i\right)^{k_{i+1}/2}\sqrt{2V_1(\bar{\varphi}_i)}}\int\frac{d\varphi}{\left(\varphi-\bar{\varphi}_i\right)^{k_i/2}} \, .
\end{equation}
Integration of the right-hand side gives power-law dependence, if $k_i>2$. Taking this observation into account, we obtain the asymptotics of the kink as $x\to-\infty$:
\begin{equation}\label{eq:power_assymptotic_left}
\varphi_{(\bar{\varphi}_i,\bar{\varphi}_{i+1})}(x) \approx \bar{\varphi}_i+\left[\frac{2}{(k_i-2)\left(\bar{\varphi}_{i+1}-\bar{\varphi}_i\right)^{k_{i+1}/2}\sqrt{2V_1(\bar{\varphi}_i)}}\right]^{2/(k_i-2)}\frac{1}{|x|^{2/(k_i-2)}} \,. 
\end{equation}
Similarly, for case of $k_{i+1}>2$, we obtain asymptotics of the kink as $x\to +\infty$:
\begin{equation}\label{eq:power_assymptotic_right}
\varphi_{(\bar{\varphi}_i,\bar{\varphi}_{i+1})}(x) \approx \bar{\varphi}_{i+1}-\left[\frac{2}{(k_{i+1}-2)\left(\bar{\varphi}_{i+1}-\bar{\varphi}_i\right)^{k_{i}/2}\sqrt{2V_1(\bar{\varphi}_{i+1})}}\right]^{2/(k_{i+1}-2)}\frac{1}{x^{2/(k_{i+1}-2)}} \,. 
\end{equation}
Below, we will use Eqs.~\eqref{eq:power_assymptotic_left} and \eqref{eq:power_assymptotic_right} to find power-law asymptotics of kinks of a $\varphi^8$ model. Note that exponential asymptotics could be found by integrating  Eq.~\eqref{eq:int_x} for the case of $k_i=2$, but a further refinement of our approximations is needed to capture the prefactor of the exponential.

\subsection{Power-law tails of the $\varphi^8$ kinks}
\label{sec:Tails}

As our featured example, we consider the triple-well $\varphi^8$ potential
\begin{equation}\label{eq:V_8}
V(\varphi)=\varphi^4(1-\varphi^2)^2.
\end{equation}
This potential has three minima: $\bar{\varphi}_1=-1$, $\bar{\varphi}_2=0$, and $\bar{\varphi}_3=1$. 
Hence, there are two kinks in the model, $\varphi_{(-1,0)}(x)$ and $\varphi_{(0,1)}(x)$, and two corresponding antikinks.

Inserting the potential \eqref{eq:V_8} into Eq.~\eqref{eq:bps} and requiring that $0<|\varphi|<1$, we have
\begin{equation}
\frac{d\varphi}{\varphi^2(1-\varphi^2)}=\sqrt{2}\:dx.
\end{equation}
Integrating this equation, we obtain (implicitly) the two kinks belonging to the topological sectors $(-1,0)$ and $(0,1)$:
\begin{equation}\label{eq:kinks_8}
2\sqrt{2}\: x = -\frac{2}{\varphi}+\ln\frac{1+\varphi}{1-\varphi} \,.
\end{equation}
The corresponding antikinks can be obtained from \eqref{eq:kinks_8} by the transformation $x\mapsto-x$:
\begin{equation}\label{eq:antikinks_8}
2\sqrt{2}\: x = \frac{2}{\varphi}-\ln\frac{1+\varphi}{1-\varphi} \,.
\end{equation}
It can be directly inferred from Eq.~(\ref{eq:antikinks_8}) that
the asymptotics of the kink $\varphi_{(0,1)}(x)$ as $x\to-\infty$ is of power-law type; likewise, for the asymptotics of the kink $\varphi_{(-1,0)}(x)$ as $x\to+\infty$. Using \eqref{eq:stenerg_bps} we obtain the energy of the static kink \eqref{eq:kinks_8}:
\begin{equation}\label{eq:M_static_kink}
M=\frac{2\sqrt{2}}{15} \,.
\end{equation}
Of course, this energy is the same for all possible kinks of the model with the potential \eqref{eq:V_8}, see also Ref.~\cite{khare}.

Now, we derive the asymptotic behavior of the kinks $\varphi_{(-1,0)}(x)$ and $\varphi_{(0,1)}(x)$. We can do this in two ways: firstly, by using the approach from Section \ref{sec:Asymptotics}, and secondly, by expanding Eq.~\eqref{eq:kinks_8} in a Taylor series around the appropriate limiting point(s). Consider the kink $\varphi_{(-1,0)}(x)$. According to the notation of Section \ref{sec:Asymptotics}, for this kink:
\begin{subequations}
\begin{alignat}{3}
\bar{\varphi}_i &= -1, \qquad k_i &&= 2,\\
\bar{\varphi}_{i+1} &= 0, \qquad k_{i+1} &&= 4,\\
V_1(\varphi) &= (1-\varphi)^2.
\end{alignat}
\end{subequations}
Equations \eqref{eq:power_assymptotic_left} and \eqref{eq:power_assymptotic_right} are approximately applicable only if $k_i$ or $k_{i+1}$ is greater than $2$, respectively. So, Eq.~\eqref{eq:power_assymptotic_right} gives power-law asymptotics of the kink:
\begin{equation}\label{eq:kink1_asymp_plus}
\varphi_{(-1,0)}(x)\approx-\frac{1}{\sqrt{2}\: x} \,,\qquad x\to +\infty.
\end{equation}
This asymptotic expression can also be obtained from Eq.~\eqref{eq:kinks_8} as shown in \cite{khare}. Indeed, the tail asymptotics emerge rather straightforwardly from the implicit kink solution as the logarithmic term becomes a small correction in the limit. At the same time, the asymptotics at $x\to -\infty$ is exponential, and it can be obtained from Eq.~\eqref{eq:kinks_8}:
\begin{equation}\label{eq:kink1_asymp_minus}
\varphi_{(-1,0)}(x)\approx-1+\frac{2}{e^2}\: e^{2\sqrt{2}\: x},\qquad x\to -\infty.
\end{equation}

For the kink $\varphi_{(0,1)}(x)$, we have:
\begin{subequations}
\begin{alignat}{3}
\bar{\varphi}_i &= 0, \qquad k_i &&= 4,\\
\bar{\varphi}_{i+1} &= 1, \qquad k_{i+1} &&= 2,\\
V_1(\varphi) &= (1+\varphi)^2.
\end{alignat}
\end{subequations}
Similarly to the previous case, from Eq.~\eqref{eq:power_assymptotic_left} we obtain power-law asymptotics at $x\to -\infty$:
\begin{equation}\label{eq:kink2_asymp_plus}
\varphi_{(0,1)}(x)\approx-\frac{1}{\sqrt{2}\: x} \,, \qquad x\to -\infty,
\end{equation}
and the exponential asymptotics at $x\to +\infty$ can be obtained from Eq.~\eqref{eq:kinks_8}:
\begin{equation}\label{eq:kink2_asymp_minus}
\varphi_{(0,1)}(x)\approx 1-\frac{2}{e^2}\: e^{-2\sqrt{2}\: x}, \qquad x\to +\infty.
\end{equation}

To summarize: in this Section, we discussed the kinks in topological sectors $(-1,0)$ and $(0,1)$ for the featured $\varphi^8$ model with the potential \eqref{eq:V_8}. In particular, we highlighted that both of these kinks have one power-law and one exponential asymptotic decay to their equilibrium background states (vacua) at $|x|\to \infty$. In what follows, we use the kink $\varphi_{(-1,0)}(x)$ and its corresponding antikink to study their their interaction via their power-law tail asymptotics. We will argue that power-law asymptotics lead to long-range interaction: a kink and an antikink ``feel'' each other at very large separations. This situation is quite different from the case of exponential tail asymptotics (for example, if $\varphi_{(-1,0)}(x)$ and its antikink were reversed in their initial configuration, or as in the classical $\varphi^4$ field theory), in which case the kink-antikink interaction is exponentially decaying.

\section{Direct Numerical Simulation of Collisions}
\label{sec:num}

The non-integrability of models such as the above-mentioned $\varphi^{2n}$ field theories suggests that exact multi-soliton solutions are not available in these systems. Nevertheless, as it is well-known from numerous prior works~\cite{csw,goodman2,dorey,weigel01,weigel02,aek01,anninos,Gani.PRE.1999,Bazeia.JPCS.2017,Bazeia.EPJC.2018,Gani.EPJC.2018}, the study of kink-antikink collisions is both particularly interesting in its own right and potentially presents a very rich phenomenology. Among the many phenomena observed in such collisions are multi-bounce windows, the fractal structure thereof, the role of the presence (or even absence~\cite{dorey}) of internal vibration modes, the ability to describe such phenomena via collective coordinate methods (or complications~\cite{weigel01,weigel02} thereof), and many others. Of critical importance to all of the above is the computational feasibility of interrogating collision phenomena via direct numerical simulations of the $(1+1)$-dimensional field theory. Indeed, the main message of the present work is that the standard approaches for setting up such simulations do \emph{not} work for higher-order field theories, such as the ones considered herein. In light of this observation, we begin by discussion the well-known $\varphi^4$ field theory. Subsequently, we compare/contrast it with the specific $\varphi^8$ model from Sec.~\ref{sec:Tails}, and finally we present some possibilities for handling the complications due to long-range interactions of kinks.

Prior to discussing the numerical results, it is relevant to briefly describe the methods used to obtain them. We discretize the governing equation of motion~\eqref{eq:nkg} on the spatial domain $x\in[-200, 200]$ with the increment $\Delta x=0.2$ (which fixes the number of Fourier modes used). We use a Fourier-based pseudospectral differentiation matrix $D_2$ as in \cite{trefethen} to approximate $\partial^2\varphi/\partial x^2$ as $D_2\varphi$ subject to periodic boundary conditions. 
This step turns the PDE into a semi-discrete system of second-order-in-time ordinary differential equations (ODEs). These are trivially rewritten as a first-order system of ODEs and integrated forward in time using MATLAB's {\tt ode45} differential equations solver with adaptive time stepping and error control.

\subsection{The standard example: $\protect\varphi^4$ field theory}
\label{sec:phi4}

The classical $\varphi^4$ field theory is determined by the potential $V(\varphi)=\frac{1}{2}(\varphi^2-1)^2$ (see, e.g., \cite{manton,vach06,aek01}). The stationary kink solution of this model in the $(-1,1)$ topological sector, i.e., the solution of the BPS equation~\eqref{eq:bps}, is $\varphi_{(-1,1)}(x)=\tanh(x)$. By Lorentz boosting the stationary solution, we obtain a traveling kink solution: 
\begin{equation}
 \varphi_v(x,t) = \varphi_{(-1,1)}\left(\frac{x-vt}{\sqrt{1-v^2}}\right) = \tanh\left(\frac{x-vt}{\sqrt{1-v^2}}\right)   
\end{equation}
for any kink velocity $v$ such that $-1<v<1$. A traveling antikink solution is given by $-\varphi_v(x,t)$ for this model. A kink moving to the right with velocity $v$, shifted to the left by the amount $x_0$, is then given by $\varphi_v(x+x_0,t)  = \varphi_{(-1,1)}\left(\frac{x+x_0-vt}{\sqrt{1-v^2}}\right)$, and an antikink moving to the left with opposite velocity, shifted to the right by the amount $x_0$, is likewise given by $-\varphi_{-v}(x-x_0,t) = -\varphi_{(-1,1)}\left(\frac{x-x_0+vt}{\sqrt{1-v^2}}\right)$.

Therefore, the function 
\begin{equation}
\begin{aligned}
\varphi(x,t) &= \varphi_v(x+x_0,t)-\varphi _{-v}(x-x_0,t)-1 \\
&=\varphi_{(-1,1)}\left(\frac{x+x_0-vt}{\sqrt{1-v^2}}\right)-\varphi_{(-1,1)}\left(\frac{x-x_0+vt}{\sqrt{1-v^2}}\right)-1
\end{aligned}
\label{ansatz1_eqn}
\end{equation}
represents a waveform consisting of a kink and an antikink with initial separation $2x_0$. The kink and antikink in this pair have equal and opposite velocities $\pm v$. Furthermore, $\varphi(x,t)$ given by Eq.~\eqref{ansatz1_eqn} is an approximate solution to the PDE~\eqref{eq:nkg}, for $x_0$ sufficiently large and $t$ sufficiently small. In fact, as long as the (midpoints of the) kink and antikink are separated by about $20$ units (for example, a stationary solution with $x_0=10$, or a traveling solution with $v=1/2$, $x_0=20$ and $0\leq t\leq 20$), $\varphi(x,t)$ from Eq.~\eqref{ansatz1_eqn} satisfies the field equation~\eqref{eq:nkg} to approximately standard machine precision, i.e., on the order of $10^{-16}$. Specifically by ``satisfies the PDE'' we mean that the residual, as measured by the value of 
\begin{equation}
    maxAbsPde(t) := \displaystyle\max_x \left|\frac{\partial^2\varphi}{\partial t^2} - \frac{\partial^2\varphi}{\partial x^2} + V'(\varphi)\right|,
    \label{eq:maxAbsPDE}
\end{equation}
is suitably small. Therefore, it is reasonable to use Eq.~\eqref{ansatz1_eqn} to generate initial conditions for kink-antikink collisions with
\begin{equation}
\begin{aligned}
\varphi(x,0) &= \varphi _{v}(x+x_0,0)-\varphi _{-v}(x-x_0,0)-1 \\
&= \varphi_{(-1,1)}\left(\frac{x+x_0}{\sqrt{1-v^2}}\right)-\varphi_{(-1,1)}\left(\frac{x-x_0}{\sqrt{1-v^2}}\right)-1,
\end{aligned}
\label{eq:PDE_IC1}
\end{equation}
and
\begin{equation}
\begin{aligned}
\frac{\partial \varphi}{\partial t}(x,0) &= \frac{\partial \varphi_v}{\partial t}(x+x_0,0) - \frac{\partial \varphi_{-v}}{\partial t}(x-x_0,0) \\
&= -\frac{v}{\sqrt{1-v^2}}\varphi_{(-1,1)}^{\prime }\left(\frac{x-x_0}{\sqrt{1-v^2}}\right)-\frac{v}{\sqrt{1-v^2}}\varphi_{(-1,1)}^{\prime }\left(\frac{x+x_0}{\sqrt{1-v^2}}\right),
\end{aligned}
\label{eq:PDE_IC2}
\end{equation}
where primes denote differentiation with respect to the function's argument. However, for separations $2x_0 \lesssim 20$ units, the value of $maxAbsPde$ (evaluated from the numerical solution of the PDE starting from the initial conditions in Eqs.~\eqref{eq:PDE_IC1} and \eqref{eq:PDE_IC2} with $v=0$) decreases exponentially with  $x_0$;  for stationary solutions, its magnitude is on the order of $10^{-7}$ at a separation of $2x_0=10$, and on the order of $10^{-3}$ at a separation of $2x_0=5$. 

It is relevant to mention here that the ansatz in Eqs.~\eqref{eq:PDE_IC1} and \eqref{eq:PDE_IC2} is suitable not only for direct numerical simulations, but also for collective coordinate approximations of the PDE dynamics~\cite{Sugiyama,csw,aek01,malomed,GaKuLi,weigel01,weigel02,baron01,berenstein,hata01,javidan,shnir,DDKCS}. In particular, one can use Eqs.~\eqref{eq:PDE_IC1} and \eqref{eq:PDE_IC2} with $x_0 - v t \mapsto X(t)$ as a new variable (the \emph{collective coordinate}) that determines the dynamic location of the kink's center. In some of the latter references, more elaborate ans{\"a}tze involving also a coordinate characterizing the kink's internal (vibration) mode were considered. However, these considerations are  beyond the scope of the present study. The principal features of a collective coordinate approach to kink-antikink interactions in the higher-order field theories of interest herein are presented in the Appendix, for completeness.

\subsection{The present case: $\varphi^8$ field theory}
\label{sec:phi8}

\begin{figure}
\centering
\includegraphics[width=.5\textwidth]{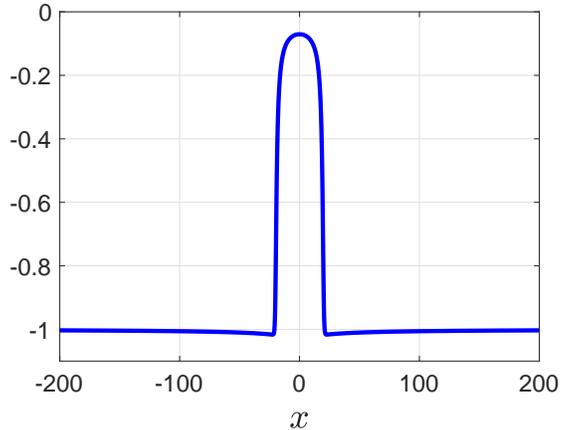}
\caption{Graph of the $\varphi^8$ kink-antikink linear superposition given in Eq.~\eqref{phi8sum_eqn} at $t=0$ with $x_0=20$.}
\label{fig:phi8sum}
\end{figure}

Now, consider the governing PDE~\eqref{eq:nkg} with the potential $V(\varphi) = \varphi^4(\varphi ^2-1)^2$ as in Sec.~\ref{sec:Tails} above. We can study a kink-antikink collision interaction by adapting the ansatz from Eq.~\eqref{ansatz1_eqn} as follows:
\begin{equation}
\varphi(x,t) = \varphi_{(-1,0)}\left(\frac{x+x_{0}-vt}{\sqrt{1-v^{2}}}\right) + \varphi_{(0,-1)}\left(\frac{x-x_{0}+vt}{\sqrt{1-v^{2}}}\right),
\label{phi8sum_eqn}
\end{equation}
where $\varphi_{(-1,0)}$ and $\varphi_{(0,-1)}$ are given implicitly by Eq.~\eqref{eq:kinks_8}. As in the $\varphi^4$ example above, we can use Eq.~\eqref{phi8sum_eqn} to generate initial conditions for a collision simulation. However, because of the power-law tails of the kink and antikink, we are presented with a problem. In Fig.~\ref{fig:phi8sum}, we show a graph of the ansatz given in Eq.~\eqref{phi8sum_eqn} for $v=0$ and $x_0=20$. Near the point at which the kink starts to rise from $\varphi=-1$ (at $x\approx-20$), one can see that the shape dips below $\varphi=-1$; there is a similar undershoot at the symmetric point on the other side of $x=0$. The left undershoot is due to power-law nature of the antikink (it is still significantly less than zero), and vice versa for the right undershoot. Also, note that this function does not get very close to $\varphi=0$ for $|x|\approx 0$. These observations imply that the kink and antikink are {\it not} sufficiently well separated for Eq.~\eqref{phi8sum_eqn} to be a suitable ansatz for a kink-antikink interaction/collision simulation. Let us now make this notion of poor approximation quantitatively precise.

To this end, consider the PDE residual $maxAbsPde$ defined in Eq.~\eqref{eq:maxAbsPDE}. Now, we substitute Eq.~\eqref{phi8sum_eqn} into Eq.~\eqref{eq:maxAbsPDE} to determine whether this ansatz provides a suitable \emph{approximate} kink-antikink solution to the PDE~\eqref{eq:nkg}. Evaluating $maxAbsPde$ numerically (keeping in mind that $\partial^2\varphi/\partial t^2=0$ for the stationary solution with $v=0$), using the pseudospectral differentiation matrix $D_2$ to approximate $\partial^2/\partial x^2$ as mentioned above, for $x_0=5$, $10$ and $20$, we obtain the values $0.72$, $0.32$ and $0.15$ respectively, all on the order of $10^{-1}$ (see Table~\ref{table:sum} and discussion below). Recall that for the $\varphi^4$ model (Sec.~\ref{sec:phi4}), with $x_0=5$ (i.e., a stationary kink-antikink pair at a separation of $10$) we found that $maxAbsPde$ was on the order of $10^{-7}$. Thus, in contrast to the $\varphi^4$ case, even for a separation as large as $40$, we find that the linear superposition (i.e., \emph{sum}) ansatz in Eq.~\eqref{phi8sum_eqn} does {\it not} provide an approximate solution to the $\varphi^8$ equation of motion in a quantitative sense. We thus warn the numerous practitioners of such numerical computations regarding the substantial obstacles to using the classical sum ans{\"a}tze to study collisional dynamics of kinks and their interactions numerically.

\begin{figure}
\centering
\subfigure[]{
\includegraphics[scale=0.4]{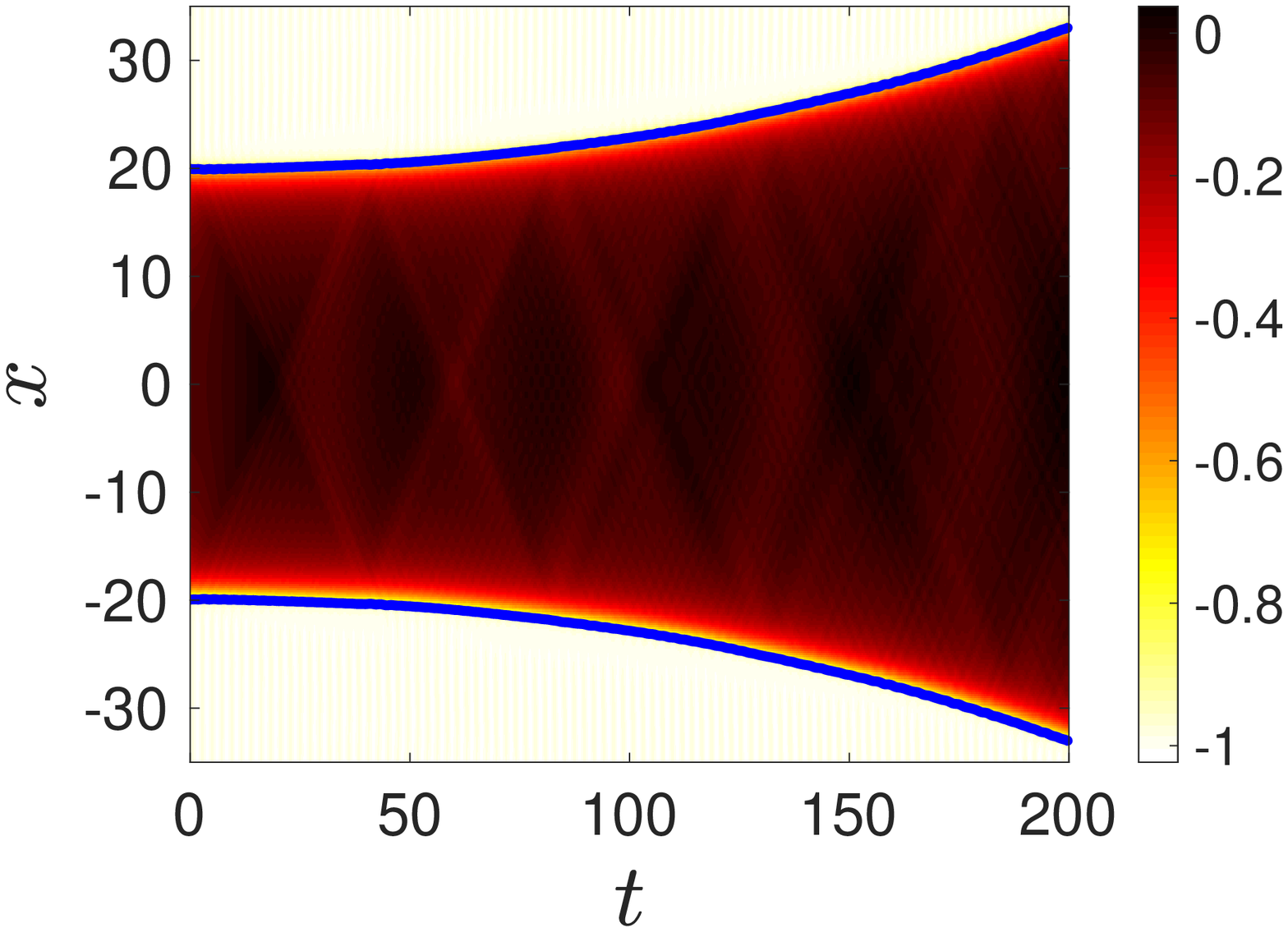}} 
\subfigure[]{
\includegraphics[scale=0.4]{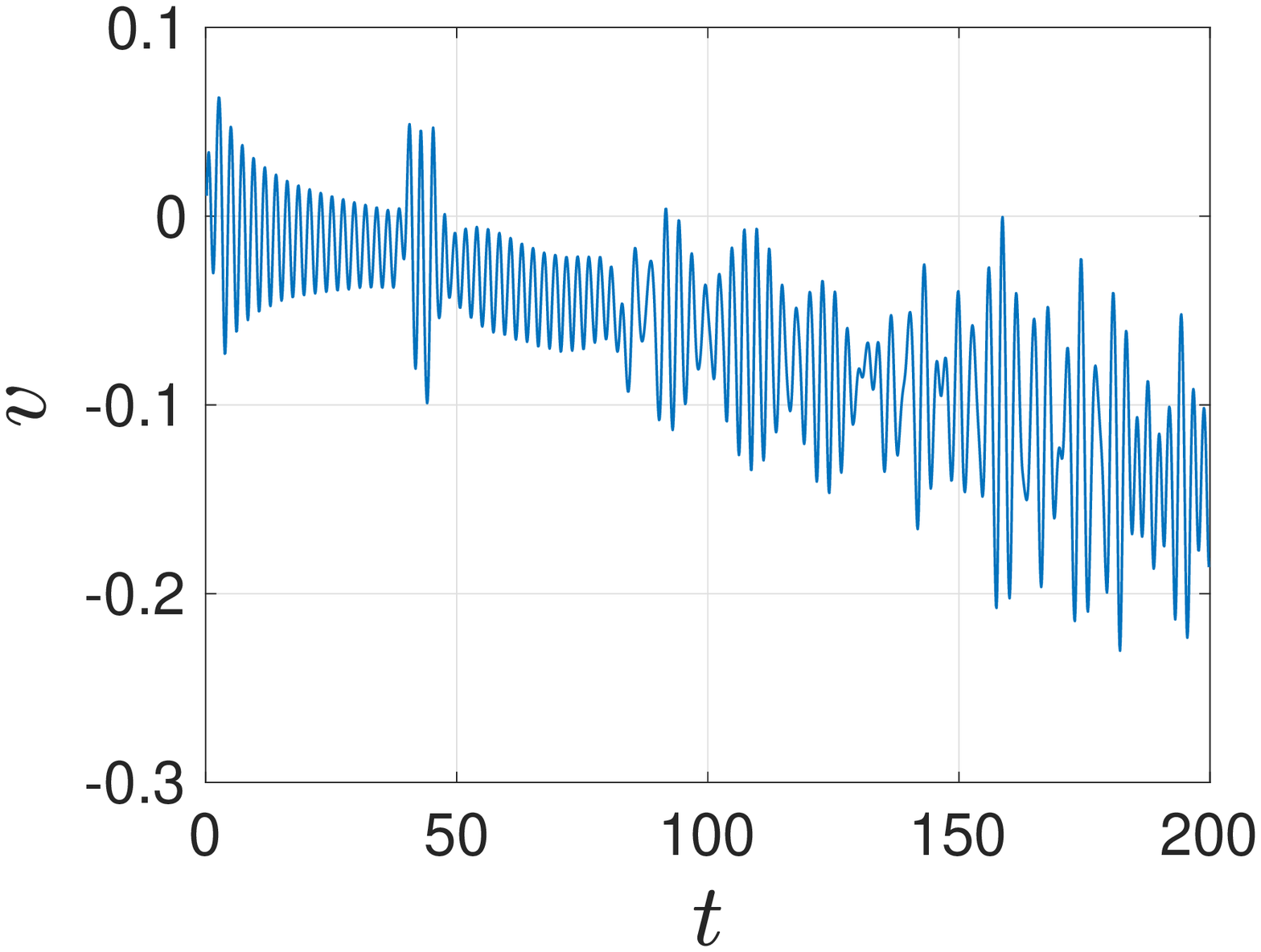}}
\caption{Using the sum ansatz, i.e., Eq.~\eqref{phi8sum_eqn}, to generate the initial conditions for the $\varphi^8$ model, we obtain (a) the contour space-time plot of the evolution of $\varphi(x,t)$ with the bottom blue curve corresponding to the kink center $x_{\mathrm{K}}$ and the top blue curve corresponding to the antikink center $x_{\mathrm{\bar{K}}}$, and (b) the plot of the velocity the kink, both from the PDE~\eqref{eq:nkg} evolution for $x_0=20$ and $v = 0$.}
\label{fig:phi8sumEvolution}
\end{figure}

Next, consider what happens when we use Eq.~\eqref{phi8sum_eqn} to create initial conditions for a prototypical kink-antikink collision simulation. We restrict ourselves to the case in which the kink and antikink are initially stationary, i.e., $v=0$; we do this to avoid the complication(s) of how an initial kinetic energy may affect the dynamics. In Fig.~\ref{fig:phi8sumEvolution}(a), we show a contour plot of the space-time evolution of the PDE solution, $\varphi(x,t)$; superimposed onto the contour plot (in this an all subsequent figures) are the curves $x=x_{\mathrm{K}}(t)$ (the location of the kink's center) and $x=x_{\mathrm{\bar{K}}}(t)$ (the location of the antikink's center). In Fig.~\ref{fig:phi8sumEvolution}(b), we show a plot of the velocity of the kink as a function of time.  Specifically, we define $x_{\mathrm{K}}$ (bottom bold curve in Fig.~\ref{fig:phi8sumEvolution}(a)) as the (approximate) intersection of $\varphi$ with $-0.83356=\varphi_{(-1,0)}(0)$. [Note that this is the $\varphi$-value of the single kink profile at $x=0$ with $x_0=0$.] The approximation is done through linear
interpolation of the two points on the shape with the $\varphi$-values closest to $-0.83356$. Similarly, in $x_{\mathrm{\bar{K}}}$ (top bold curve in Fig.~\ref{fig:phi8sumEvolution}(a)) denotes the approximate intersection of $\varphi$ with $-0.83356=\varphi_{(0,-1)}(0)$. The velocity of the kink is calculated by using the formula $v(t_i)=[x_{\mathrm{K}}(t_{i+1})-x_{\mathrm{K}}(t_{i-1})]/(t_{i+1}-t_{i-1})$, given the solution at three discrete time values $t_{i-1}$, $t_{i}$ and $t_{i+1}$ for any $i$.

Earlier work~\cite{Belendryasova.arXiv.08.2017} has suggested that a repulsive force might exists between the example kink and antikink considered above. Yet, we argue that this apparent repulsion is a result of the ansatz selected in Eq.~\eqref{phi8sum_eqn} being a poor quantitative approximation of a kink-antikink solution. Recall the undershoot below $\varphi=-1$ near $x=\pm 20$ in Fig.~\ref{fig:phi8sum}. We can think of this undershoot as providing a kind of initial ``spring board,'' which pushes the ``points'' just above the spring board upward, and consequently causes the kink to move to the left and the antikink to move to the right. Furthermore, this upward motion creates disturbances at $x\approx\pm 20$ that move upwards and then along the top of the kink-antikink combination until they meet at $x=0$. One can see these effects clearly in the contour plot in Fig.~\ref{fig:phi8sumEvolution}(a). One can also see that, after meeting at $x=0$, these disturbances are trapped between the centers of the kink and antikink, which they reach again just before $t=50$. From the plot of the kink velocity in Fig.~\ref{fig:phi8sumEvolution}(b), it is clear that just at this time the arrival of the disturbance gives another boost to the velocity of the kink in the negative direction. Again, the disturbances are reflected back towards $x=0$ and out again to the centers of the kink and antikink for another boost to the velocities, pushing them apart faster. This phenomenology holds for  other values of the initial half-separation $x_0$, down to $x_0=2.5$.

Thus, we have accounted for the apparent repulsive force, in the long-range interaction between a $\varphi^8$ kink and antikink, by showing that this ``force'' is simply the result of initial conditions that have been derived from an inaccurate sum ansatz. More specifically, we have quantified how this ansatz does not lead to a sufficiently accurate description of the motion of a kink-antikink pair. An additional (albeit weaker) effect along this vein consists of the radiative wavepackets, emitted from each kink, that affect both of them in the process.

\subsection{Improved initial conditions for simulating kinks with long-range interactions}
\label{sec:improved_ic}

A steady-state solution of Eq.~\eqref{eq:nkg} satisfies $-\partial^2\varphi/\partial x^2 + V^\prime (\varphi)=0$. The last equation can be discretized as $-D_2\varphi+V^\prime(\varphi)$,  where $D_2$ is again the pseudospectral differentiation matrix as in \cite{trefethen}, on $N$ discrete and equally spaced $x$ and $\varphi$ values. Furthermore, we want the initial positions of the two topological solitons (given by $-x_0$ and $x_0$) to have specified values, which adds two more discrete equations (for a total of $N+2$). These two additional equations are $\varphi(-x_0)-\tilde{\varphi}=0$ and $\varphi(x_0)-\tilde{\varphi}=0$, where $\tilde{\varphi}$ is the $\varphi$ value of a single kink or antikink at $x=0$. The resulting set of equations is over-determined and has no solution.

As a remedy, we propose to improve the initial conditions introduced in the previous section by employing Eq.~\eqref{phi8sum_eqn} as the initializer for a weighted nonlinear least-squares minimization of the objective function 
\begin{equation}\label{eq:min_funct}
    \mathcal{I}[\varphi] = \left\Vert-D_2\varphi+V^\prime(\varphi)\right\Vert_2^2 + C\left|\varphi(-x_0)-\tilde{\varphi}\right|^2 + C\left|\varphi(x_0)-\tilde{\varphi}\right|^2,
\end{equation} 
where $\|\cdot\|_2$ is the usual Euclidean norm, and $C$ is an empirical constant. Then, we can use the minimizer $\varphi_{\min}(x)$ of $\mathcal{I}$ as the initial condition for a direct numerical simulation of kink-antikink collisions, ensuring that our initial condition quantitatively satisfies the PDE to some preset accuracy. We take $C=50$, which is sufficient to keep the initial kink and antikink locations nearly fixed at $\pm x_0$ during the minimization process. The optimization problem is solved using MATLAB's optimization toolkit, specifically via the {\tt lsqnonlin} subroutine.

\begin{figure}
\centering
\includegraphics[width=.5\textwidth]{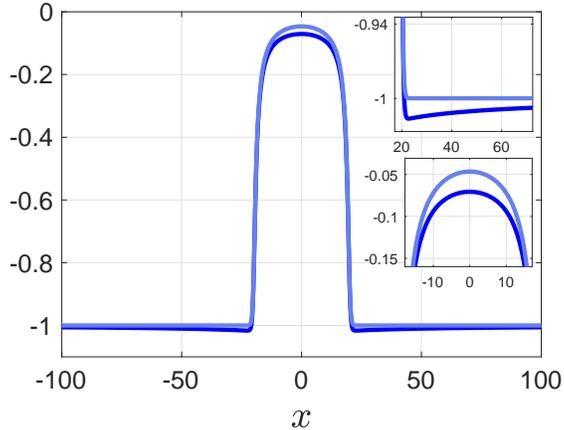}
\caption{Graphs of the unminimized sum ansatz $\varphi = \varphi_{(-1,0)}(x+20)+\varphi _{(0,-1)}(x-20)$ (light blue) and the $maxAbsPde$-minimized ansatz $\varphi = \varphi_{\min}(x)$ (dark blue). The insets show a zoom of the solution near $x=0$ and near $\varphi=-1$.}
\label{fig:phi8minimized}
\end{figure}

In Fig.~\ref{fig:phi8minimized}, for $x_0=20$, we compare the sum ansatz from Eq.~\eqref{phi8sum_eqn} (in dark blue) with the minimizer $\varphi_{\min }(x)$ of $\mathcal{I}$ (light blue). Specifically, note that we no longer observe the undershoot below $\varphi=-1$ in the plot of the minimized function. Additionally, $\varphi_{\min }(x)$ comes closer to $\varphi=0$ near $x=0$. Table \ref{table:sum} shows a more detailed quantitative comparison of $maxAbsPde$ for the minimized and (non-minimized) sum ans\"{a}tze. Specifically, the maximum value of $maxAbsPde$ when taking $\varphi=\varphi_{\min}(x)$ is several \emph{orders of magnitude} smaller than when using the sum ansatz from Eq.~\eqref{phi8sum_eqn}. Thus, we conjecture that the initial conditions generated from $\varphi_{\min}(x)$ will more accurately reflect the actual kink-antikink solution of this non-integrable field theory, at least considerably better than the non-minimized sum ansatz from Eq.~\eqref{phi8sum_eqn}. 

\begin{table}
\begin{center}
\begin{tabular}{c@{\hskip 12pt} c@{\hskip 12pt} c}
\hline
\hline
half-separation $x_0$ & non-minimized & minimized\\ 
\hline
100 & 0.029 & $1.4\times 10^{-8}$\\
\hline
50 & 0.058 & $2.2\times 10^{-7}$\\
\hline
20 & 0.15 & $8.7\times 10^{-6}$\\
\hline
10 & 0.32 & $1.4\times 10^{-4}$\\
\hline
5 & 0.72 & $2.5\times 10^{-3}$ \\
\hline
\hline
\end{tabular}
\end{center}
\caption{$absPde$ for the non-minimized and minimized sum ans\"{a}tze for the PDE initial condition.}
\label{table:sum}
\end{table}

\begin{figure}
\centering
\subfigure[]{
\includegraphics[scale=0.4]{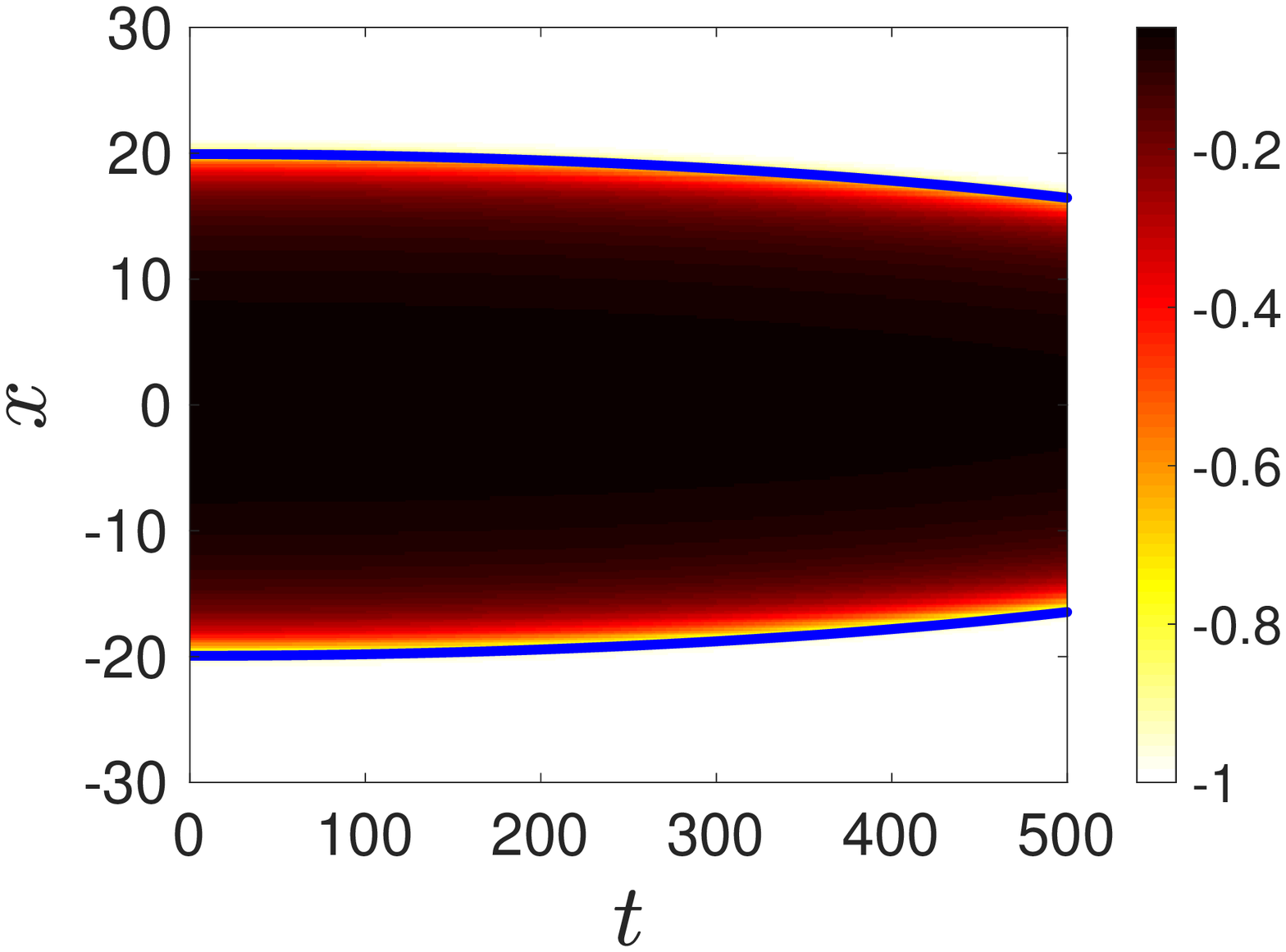}} 
\subfigure[]{
\includegraphics[scale=0.4]{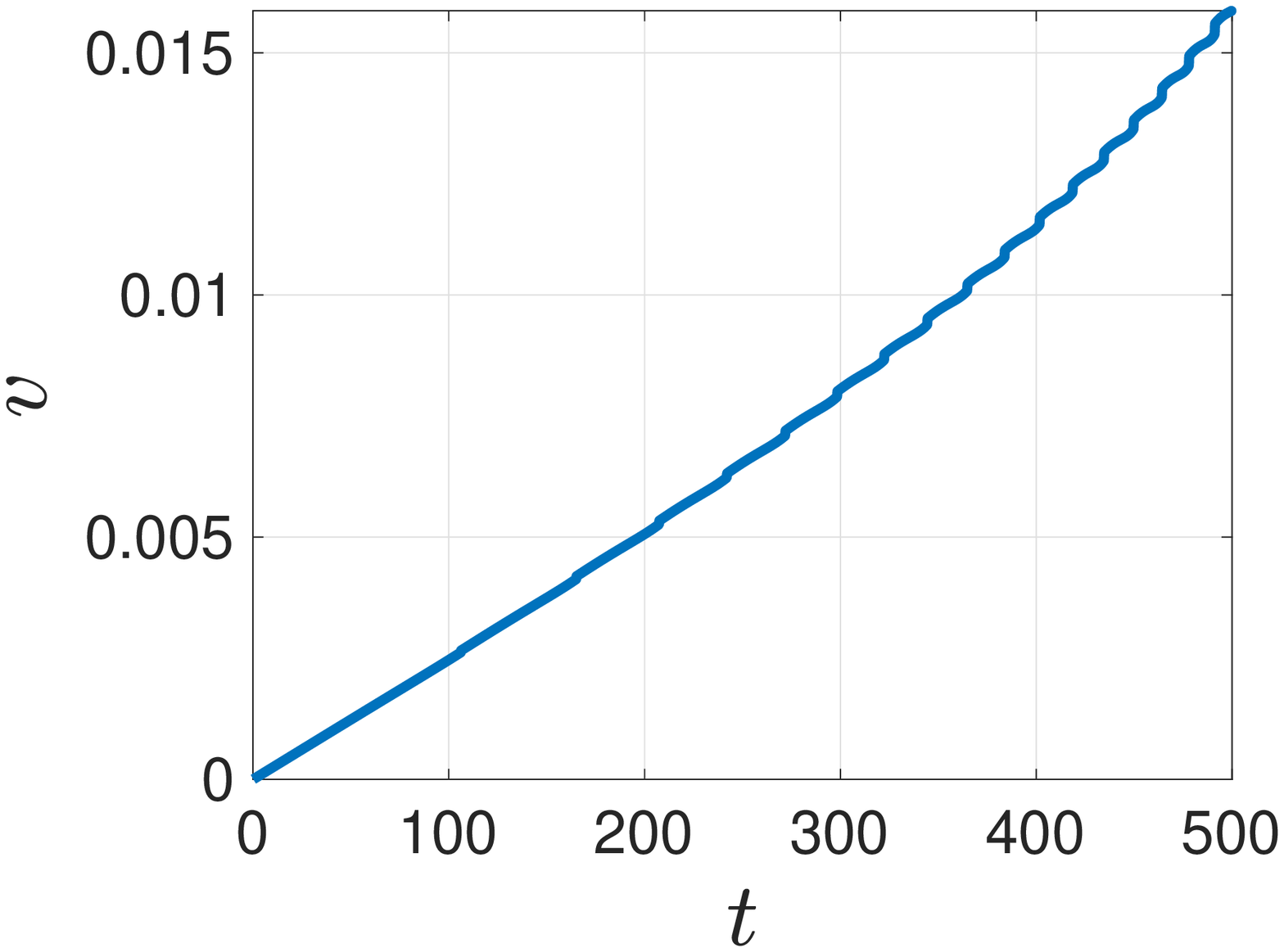}}
\caption{Using the minimizer of Eq.~\eqref{eq:min_funct} with the sum ansatz from Eq.~\eqref{phi8sum_eqn} as an initial guess for the optimization, to generate the initial conditions for the $\varphi^8$ model, we obtain (a) the contour space-time plot of the evolution of $\varphi$, and (b) the plot of the velocity, both computed from the PDE evolution, for $x_0=20$ and $v = 0$.}
\label{fig:phi8sumMinimizedEvolution}
\end{figure}

To support our conjecture, Fig.~\ref{fig:phi8sumMinimizedEvolution} shows the result of a direct numerical simulation of the PDE~\eqref{eq:nkg}, for our $\varphi^8$ model, using the minimized function $\varphi_{\min}(x)$ to generate the initial conditions. As before, this simulation corresponds to a kink-antikink interaction with $v=0$ because $\varphi_{\min}(x)$ approximates a \emph{stationary} solution of the PDE. As in Fig.~\ref{fig:phi8sumEvolution}, we show both a contour space-time plot and a velocity plot in Fig.~\ref{fig:phi8sumMinimizedEvolution}. Now, we observe an attractive force between the kink and antikink. Also, there are no
visibly discernible small disturbances moving back and forth between $x=\pm 20$. This example simulation, along with numerous other similar simulations for different $x_0$ values, suggest that we have eliminated the significant detrimental effects of the algebraic kink and antikink tails (and of radiation), and we can thus now observe the proper (effective) inter-particle interaction between the kink and antikink. Also, note that this interaction is more in line with what one would naturally expect from a PDE of the type in Eq.~\eqref{eq:nkg}, in that the $\partial^2\varphi/\partial x^2$ term tends to ``pull points down'' (pulling the kink and antikink together) when the function is concave down. Finally, the minimized ansatz $\varphi_{\min}(x)$ serves not only as an initial condition for the PDE simulations, but also as an appropriate ansatz for the collective coordinates approach described in the Appendix. 

\subsection{Other possible ans{\"a}tze}
\label{sec:other_ic}

Besides finding $\varphi$ such that $\mathcal{I}$ in Eq.~\eqref{eq:min_funct} is minimized, there are other possible setups that could be used to generate initial conditions for direct numerical simulation of the PDE~\eqref{eq:nkg} (and possibly for collective coordinate approaches as well). For example, one can use a \emph{product} (rather than a sum) ansatz. Such an ansatz, customized for our $\varphi^8$ model, is
\begin{equation}\label{eq:prod_ans}
\varphi(x,t) = \left[ \varphi_{(-1,0)}\left(\frac{x+x_{0}-vt}{\sqrt{1-v^{2}}}\right)+1\right]\left[\varphi_{(0,-1)}\left(\frac{x-x_{0}+vt}{\sqrt{1-v^{2}}}\right)+1\right] - 1,
\end{equation}
which we term the \emph{product ansatz}. 

\begin{figure}
\centering
\subfigure[]{
\includegraphics[scale=0.4]{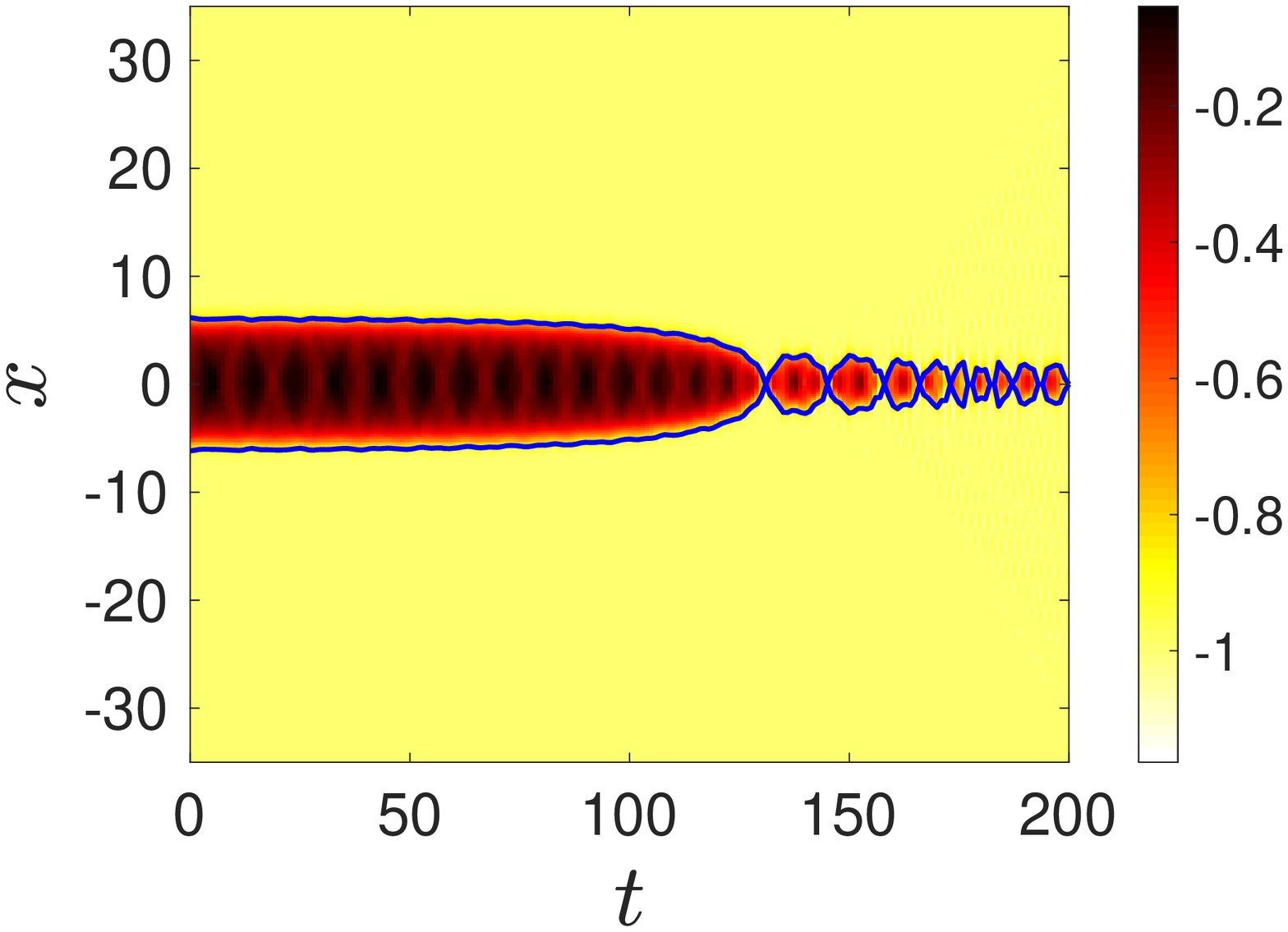}} 
\subfigure[]{
\includegraphics[scale=0.4]{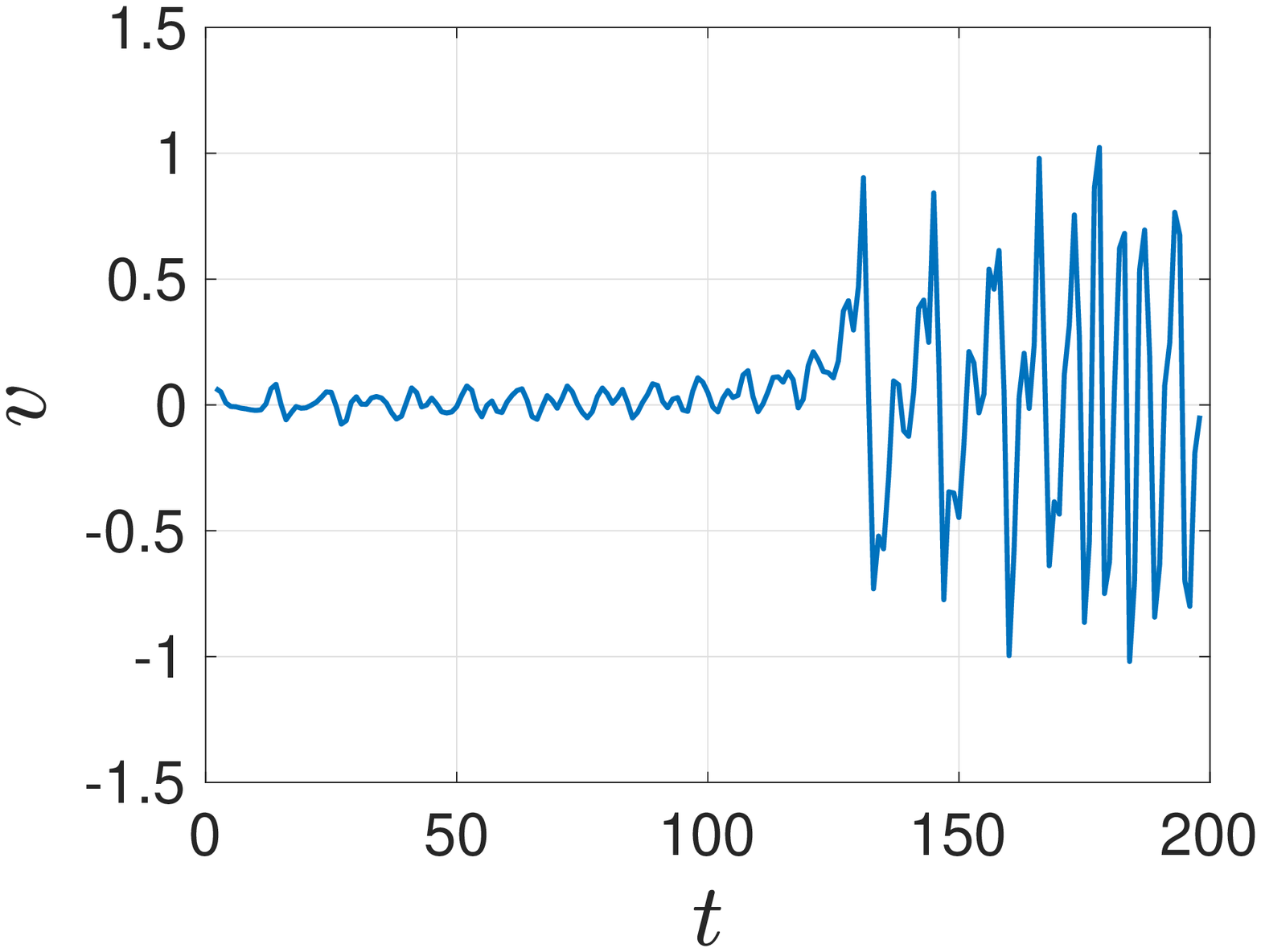}}
\subfigure[]{
\includegraphics[scale=0.4]{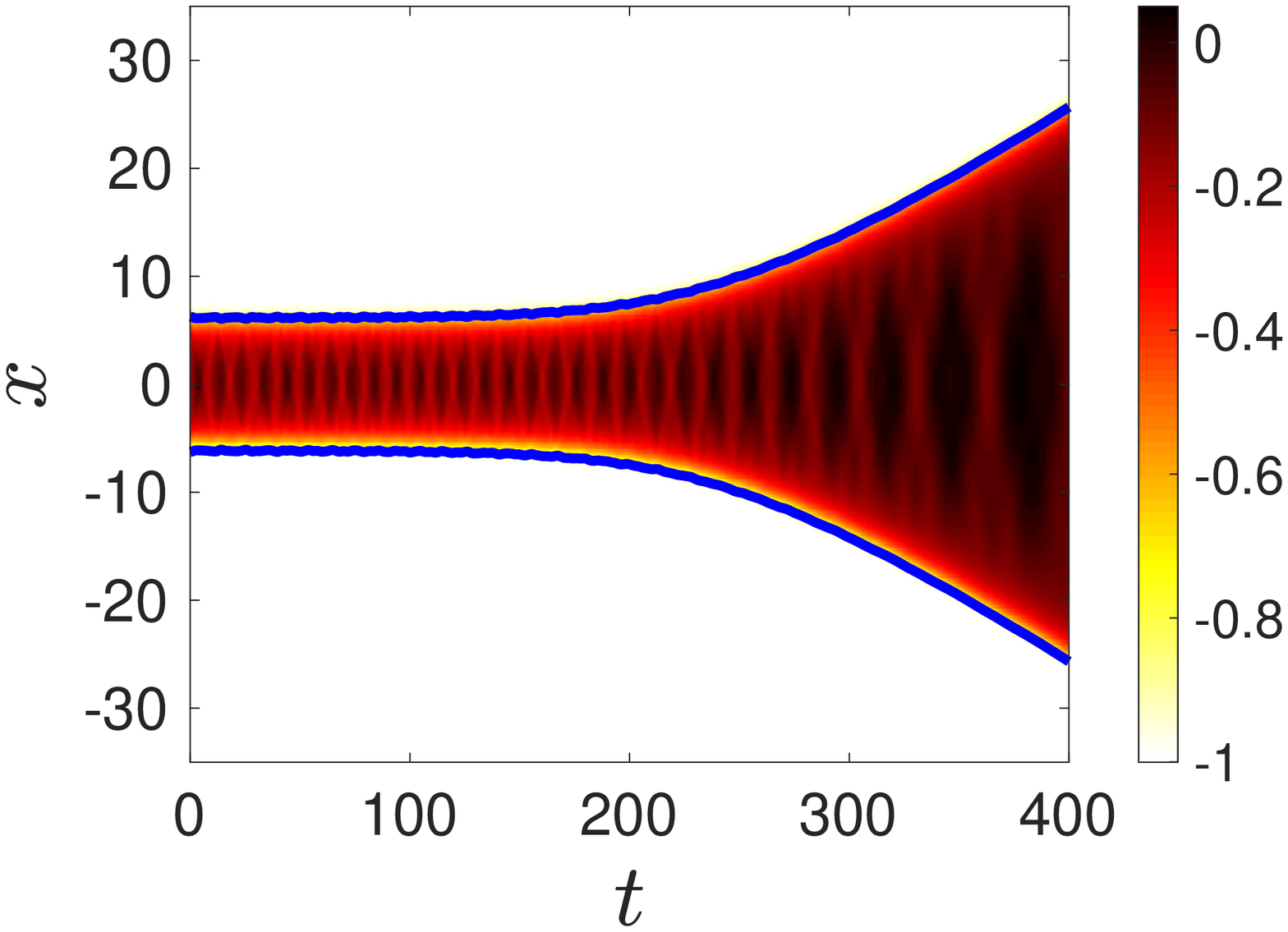}} 
\subfigure[]{
\includegraphics[scale=0.4]{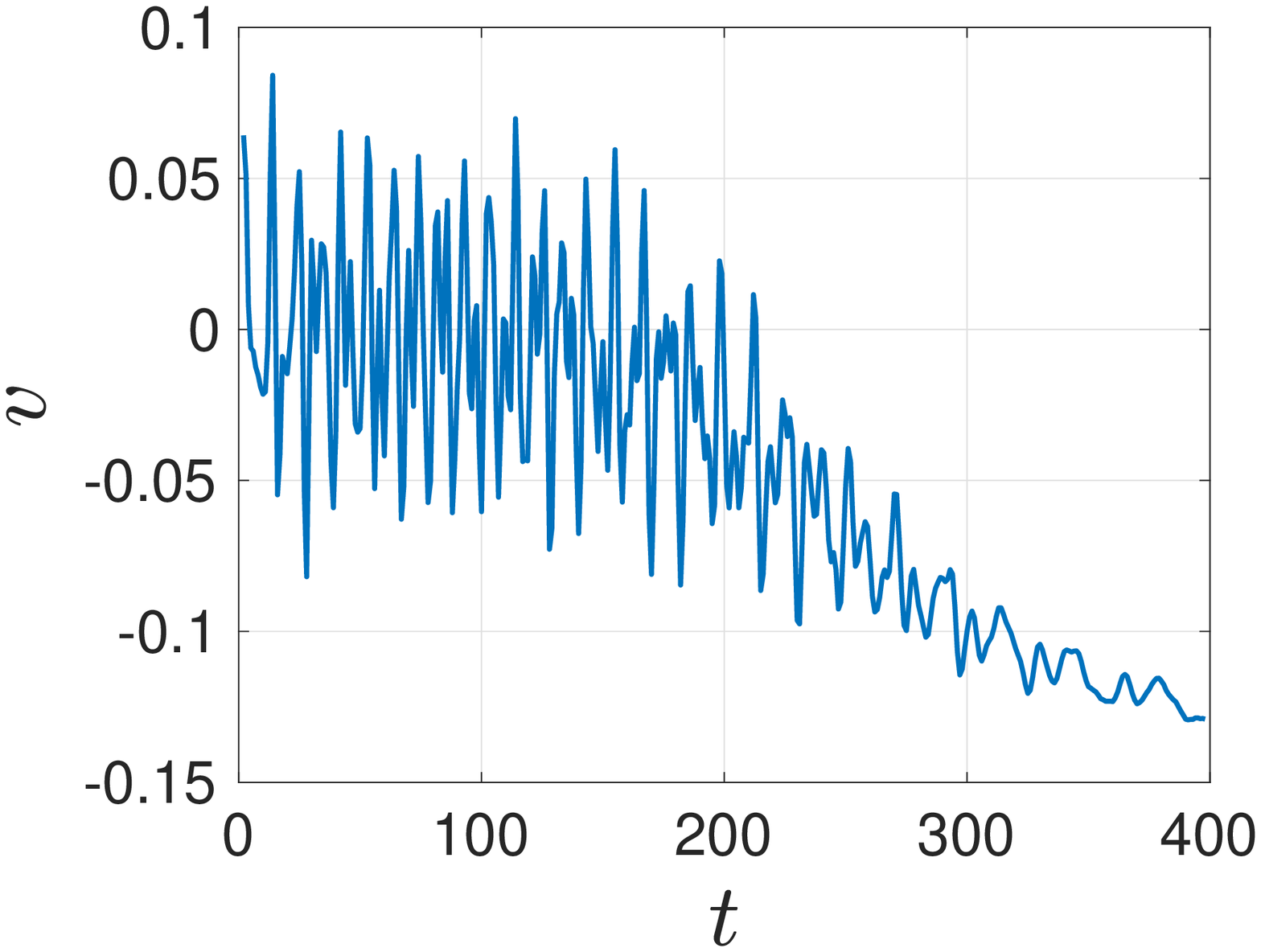}}
\caption{Using the product ansatz from Eq.~\eqref{eq:prod_ans} (not minimized) to generate the initial conditions for the $\varphi^8$ model, we obtain (a) the contour spate-time plot and (b) the velocity plot of PDE evolution for $x_0=6.2$ ($v = 0$), and (c) the contour spate-time plot and (d) the velocity plot of PDE evolution for $x_0=6.3$ ($v = 0$).}
\label{fig:product_critical}
\end{figure}

\begin{figure}
\centering
\includegraphics[width=.5\textwidth]{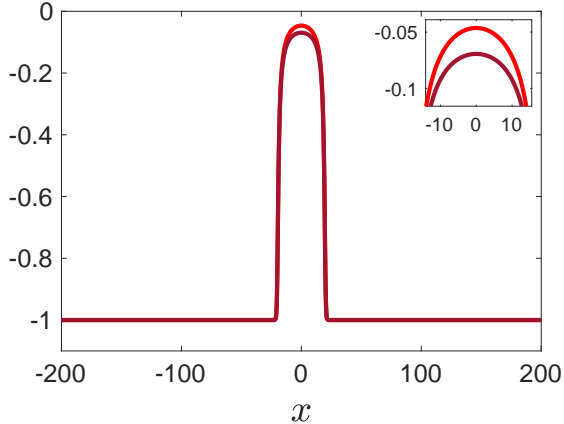}
\caption{Graph comparing the product ansatz from Eq.~\eqref{eq:prod_ans} (dark red) and its minimized counterpart (light red).}
\label{fig:productMinVsNon}
\end{figure}

\begin{figure}
\centering
\subfigure[]{
\includegraphics[scale=0.4]{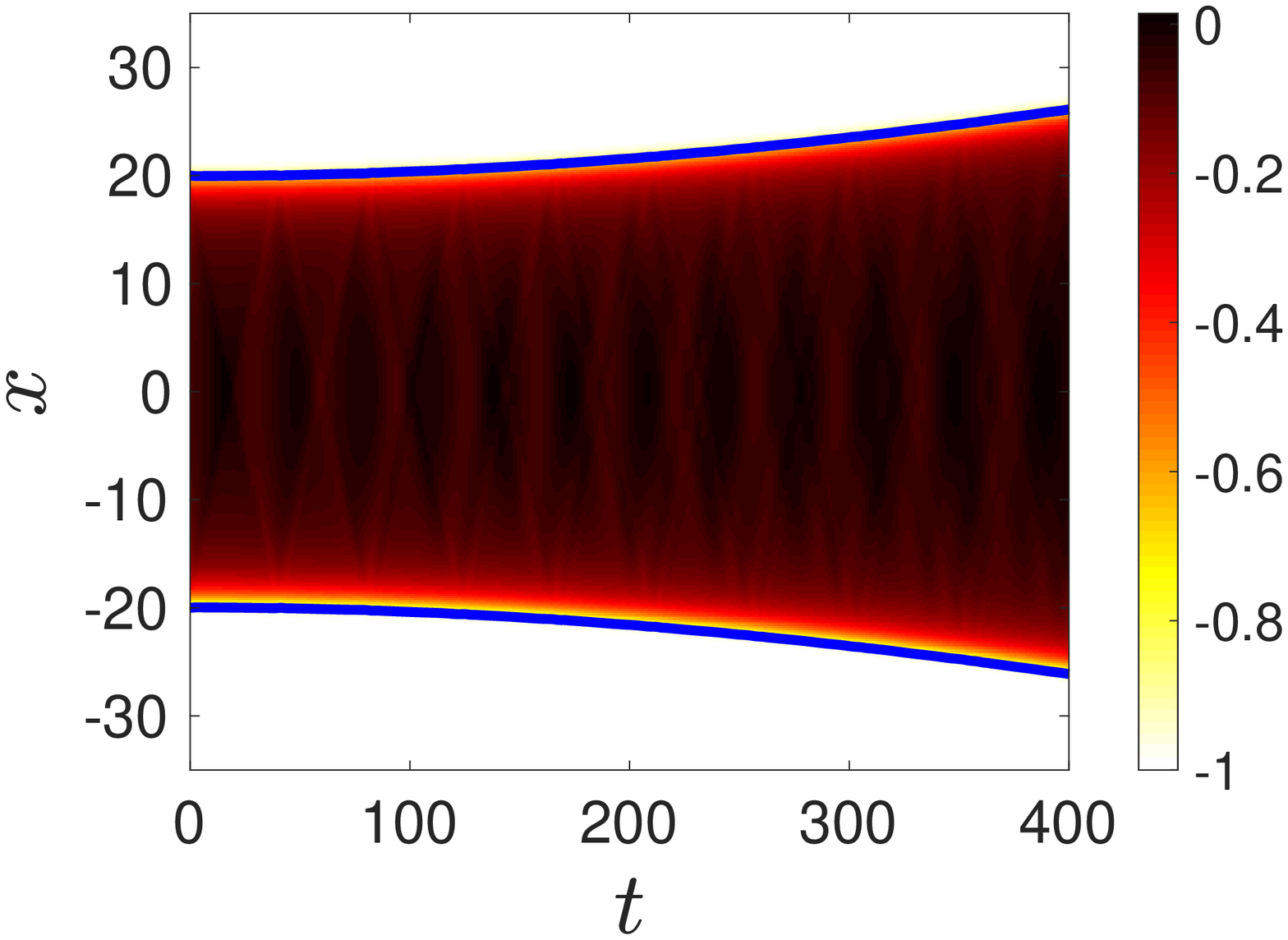}} 
\subfigure[]{
\includegraphics[scale=0.4]{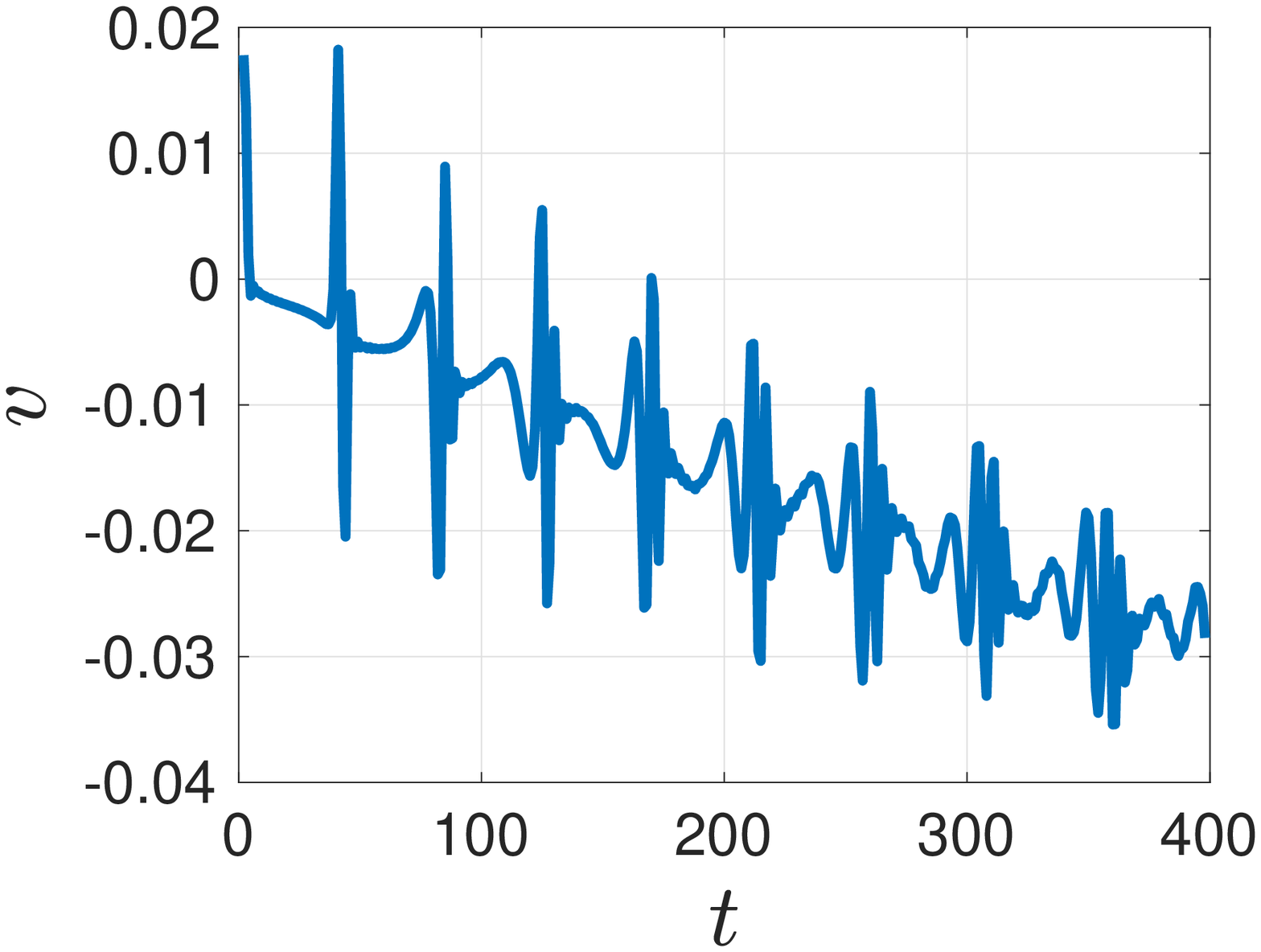}}
\subfigure[]{
\includegraphics[scale=0.4]{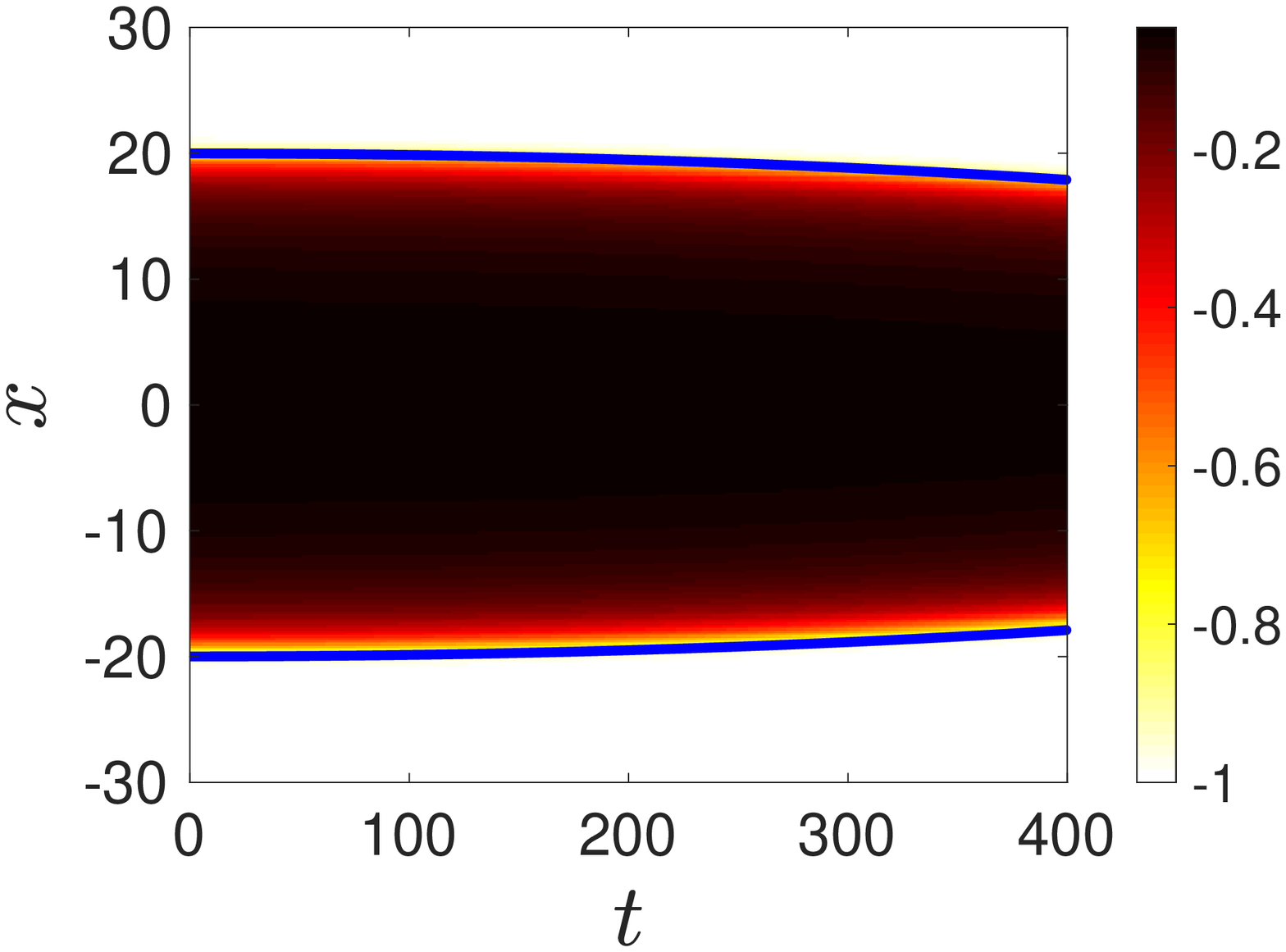}} 
\subfigure[]{
\includegraphics[scale=0.4]{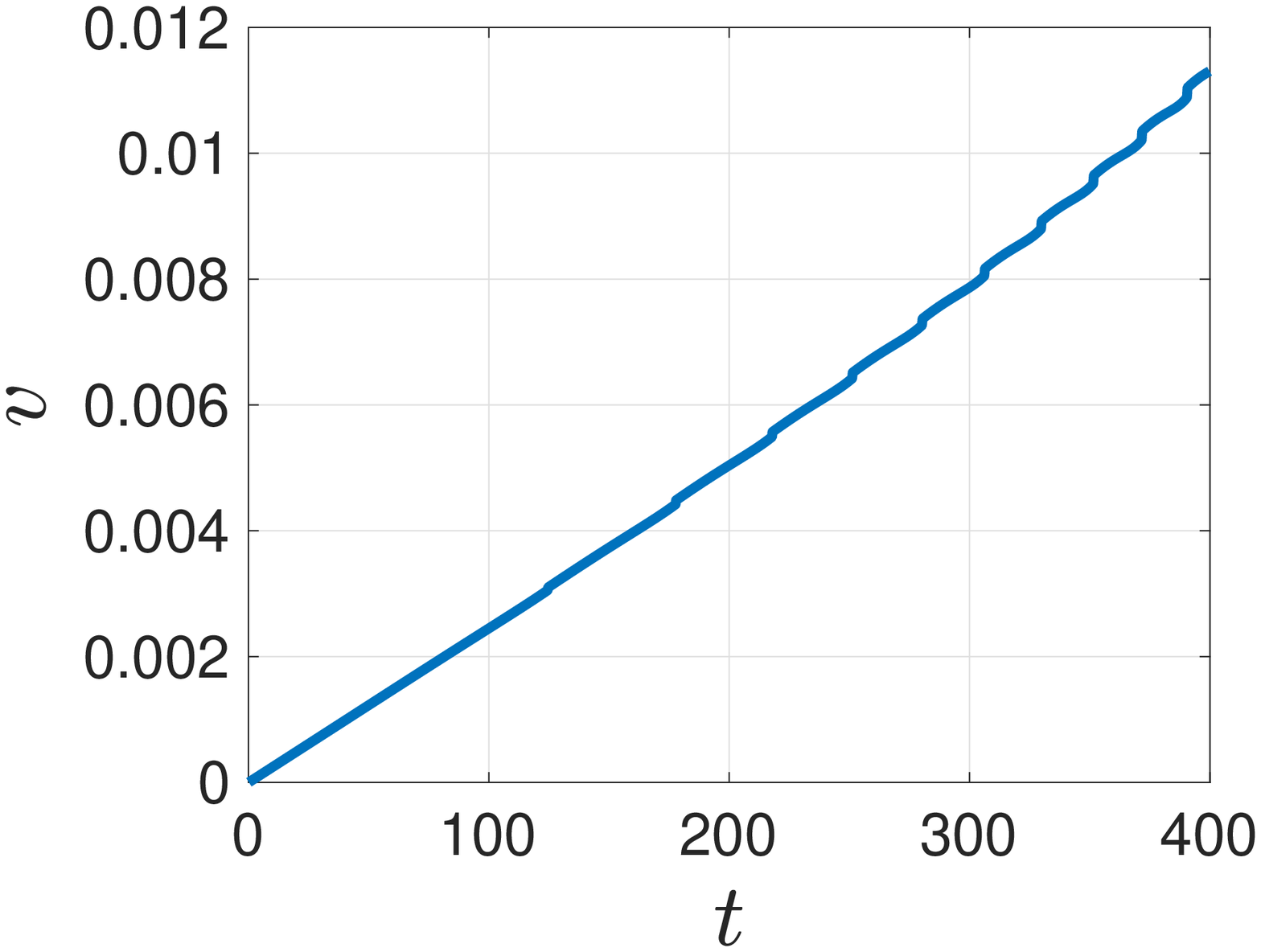}}
\caption{Using the product ansatz from Eq.~\eqref{eq:prod_ans} (not miminized) to generate the initial conditions for the $\varphi^8$ model, we obtain (a) the contour space-time plot and (b) the velocity plot, while using minimization of the product ansatz, we obtain the corresponding (c) contour space-time plot and (d) velocity plot; all panels are for $x_0=20$, $v = 0$.}
\label{fig:productNonMinimized}
\end{figure}

Numerical simulations starting from Eq.~\eqref{eq:prod_ans} as an initial condition (again with $v=0$), once again exhibit a repulsion  initially (though, weaker than for the sum ansatz) for $x_0>x_c$ where $6.2<x_c<6.3$.  For $x_0<x_c$, attraction is found in the direct numerical simulations. Figure \ref{fig:product_critical} shows the contour and the velocity plots for $x_0=6.2$, which leads to attraction, and for $x_0=6.3$, which leads to repulsion, having used the product ansatz in Eq.~\eqref{eq:prod_ans} to generate the initial conditions. In other words, the product ansatz creates the illusion of a possible saddle point configuration near $x_0=x_c$, such that attraction ensues for smaller (and repulsion for larger) initial sperations between the kink and the antikink. Minimization can be applied to the product ansatz from Eq.~\eqref{eq:prod_ans} as well, along the lines of Sec.~\ref{sec:improved_ic}. The results are similar to those corresponding to using the minimized sum ansatz; in fact, the resulting minimized functions are nearly identical. Naturally, the output of this procedure decreases the undershoot from $-1$ (admittedly weaker with the product ansatz than with the sum ansatz) and passes even closer to $\varphi=0$ in the vicinity of $x=0$; see Fig.~\ref{fig:productMinVsNon}. Minimizing the product ansatz also leads to generic attraction, as we have come to expect, at this point, from the $\varphi^8$ field theory. All of these observations are illustrated in Fig.~\ref{fig:productNonMinimized}.

\begin{figure}[tbp]
\centering
\subfigure[]{
\includegraphics[scale=0.4]{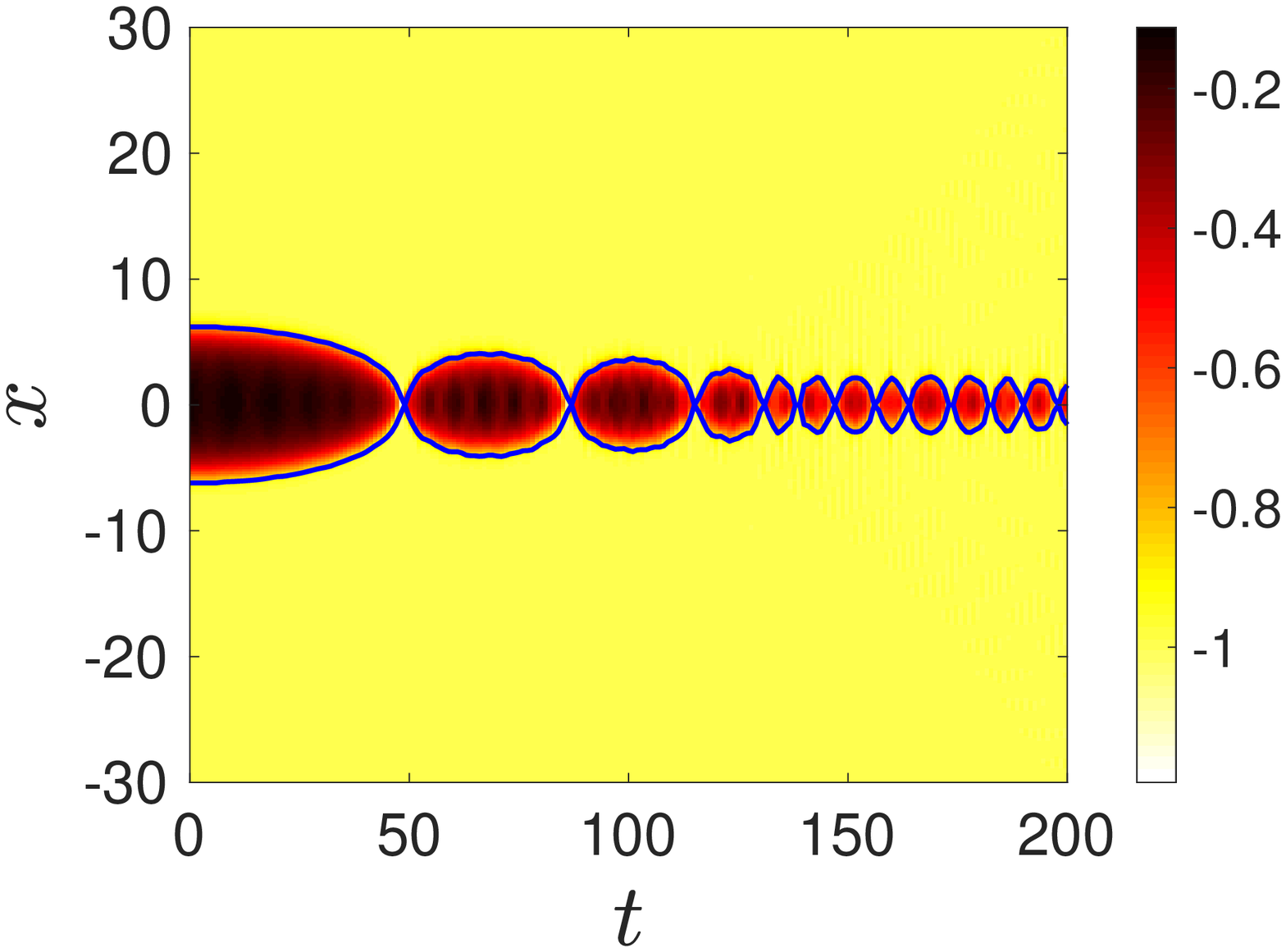}} 
\subfigure[]{
\includegraphics[scale=0.4]{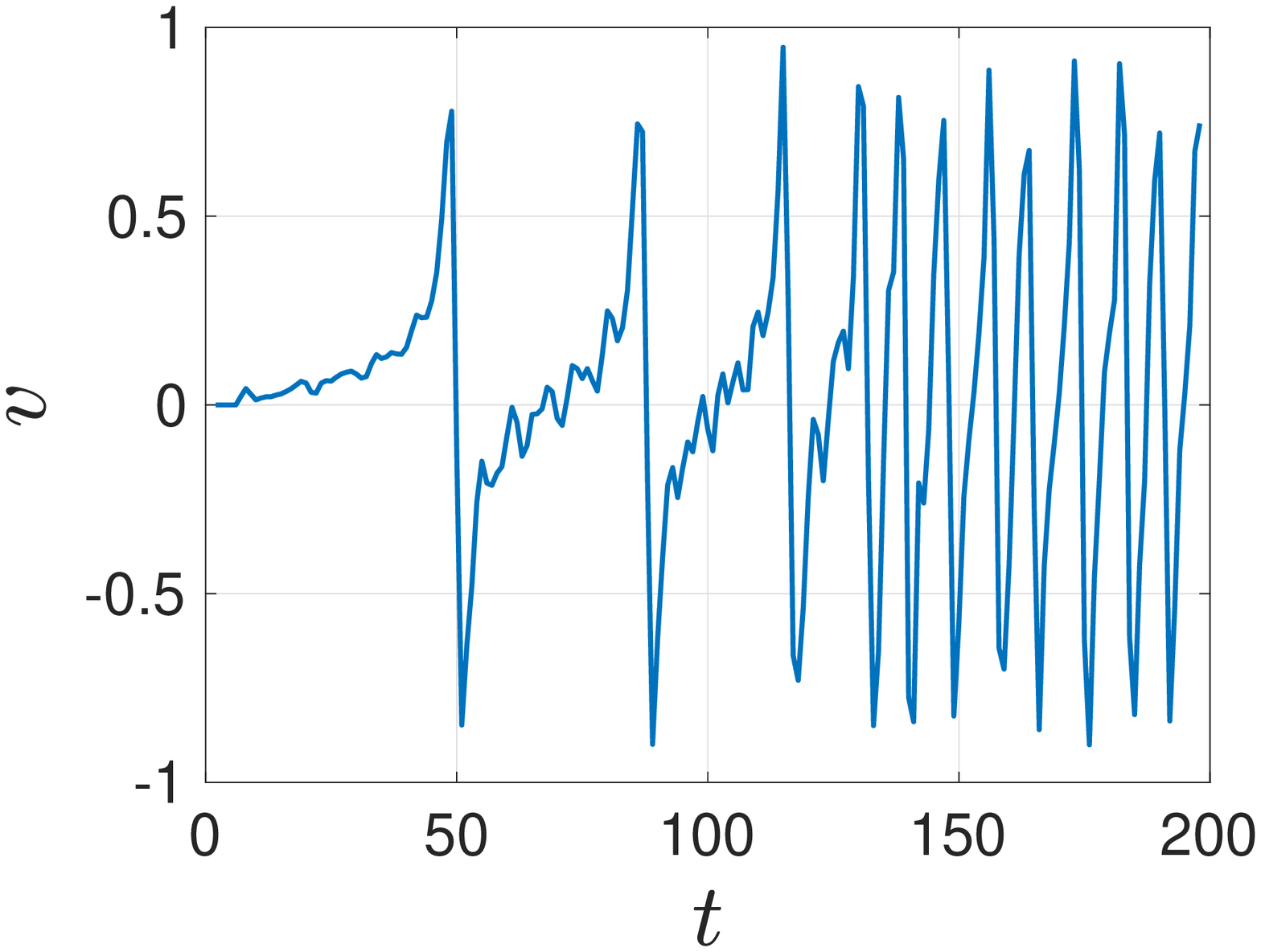}}
\subfigure[]{
\includegraphics[scale=0.4]{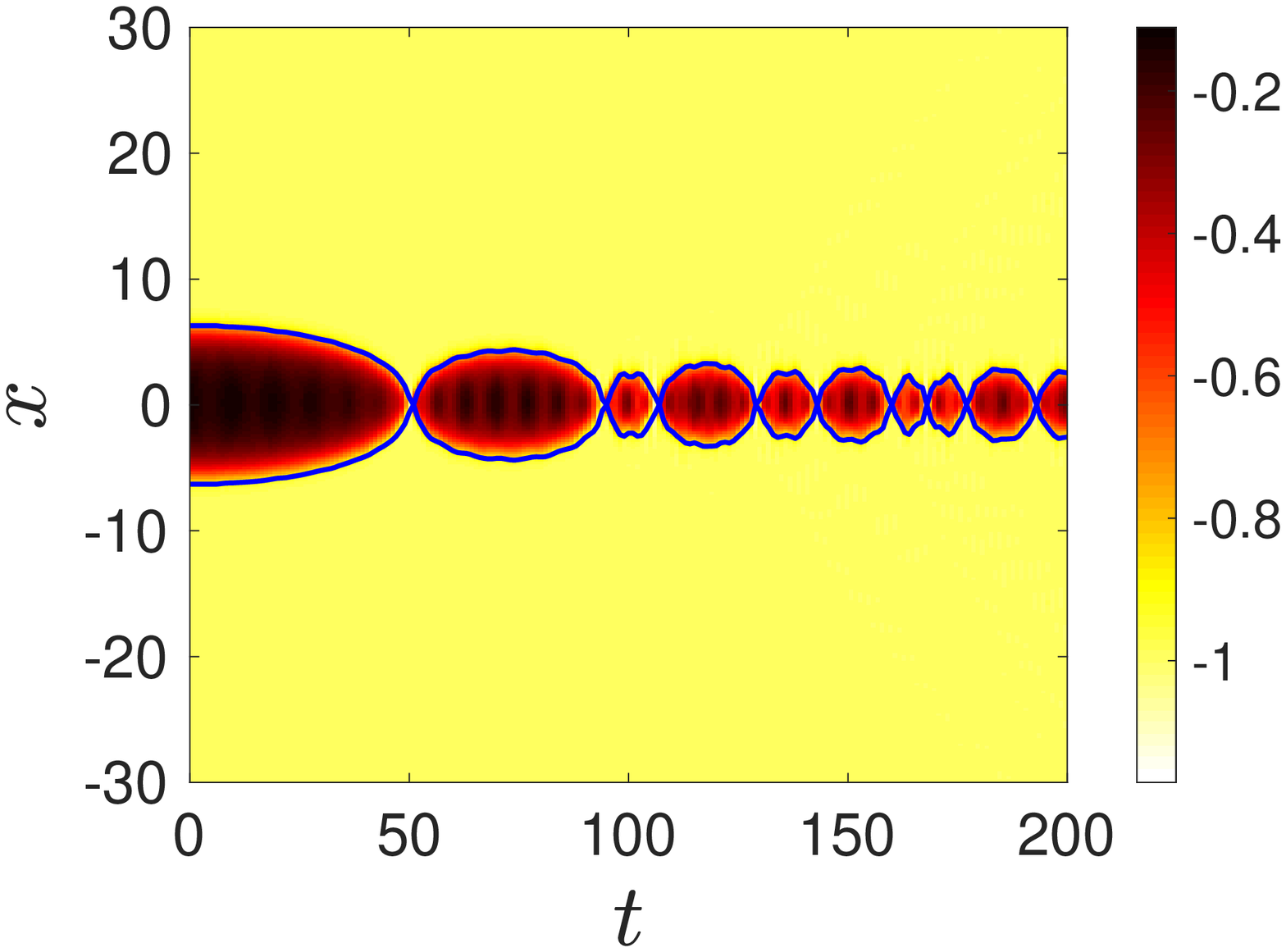}} 
\subfigure[]{
\includegraphics[scale=0.4]{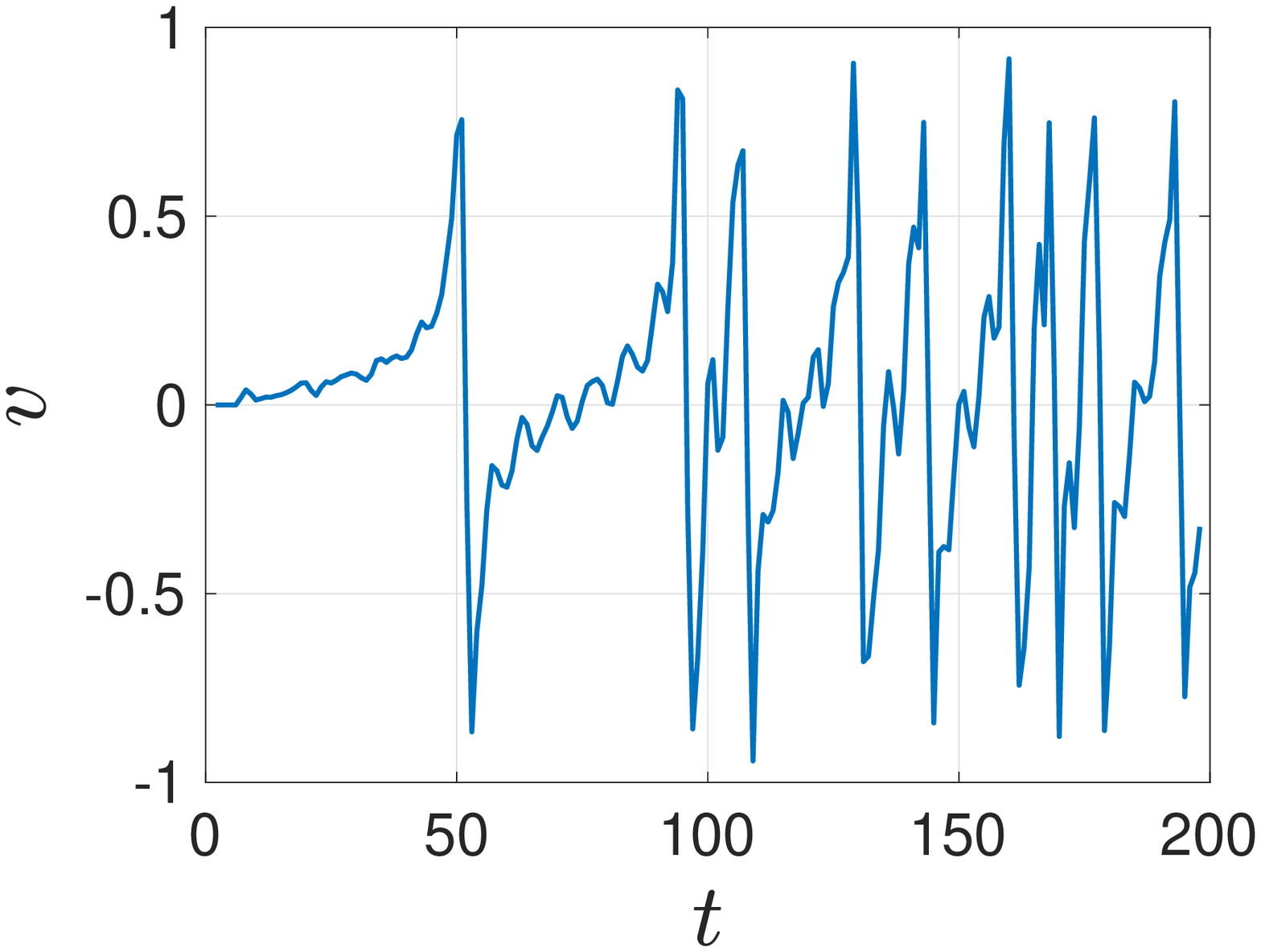}}
\caption{Using the split-domain ansatz from Eq.~\eqref{eq:phi8_split} (not minimized) to generate the the initial conditions for the $\varphi^8$ model, we obtain (a) the space-time contour plot of $\varphi$ and (b) the kink velocity plot, both from the PDE evolution, for $x_0=6.2$ and  $v = 0$. Meanwhile, (c) the space-time contour plot of $\varphi$ and (d) the kink velocity plot correspond to the PDE evolution for $x_0=6.3$ and $v = 0$.}
\label{fig:splitdomain}
\end{figure}

A third option is to treat the kink and antikink ``completely separately,'' meaning to use the kink formula for $x<0$ and the antikink formula for $x\geq 0$. To accomplish such a feat, we define
\begin{equation}\label{eq:phi8_split}
\varphi(x,t)=[1-H(x)] \varphi_{(-1,0)}\left(\frac{x+x_{0}-vt}{\sqrt{1-v^{2}}}\right)+H(x) \varphi_{(0,-1)}\left(\frac{x-x_{0}+vt}{\sqrt{1-v^{2}}}\right),
\end{equation}
which we term the \emph{split-domain ansatz}. Here, $H(x)$ is the Heaviside unit-step function. Using this ansatz to generate the initial conditions for a PDE simulation, we plot the contours of $\varphi$ and the kink velocity for $x_0=6.2$ and $x_0=6.3$ in Fig.~\ref{fig:splitdomain}. Contrary to what was the case for the product ansatz, we observe attraction for both $x_0$ values. It is perhaps natural to expect that the split-domain ansatz is the most accurate unminimized one (i.e., among the more standard ones that have not been ``optimized'' via our proposed minimization procedure), but it is expected to be limited in accuracy in the vicinity of $x=0$ due to the derivative discontinuity introduced in Eq.~\eqref{eq:phi8_split}. This observation is substantiated by the kink-antikink dynamics shown in Fig.~\ref{fig:splitdomain}.

\begin{figure}[ht]
\centering
\includegraphics[width=.5\textwidth]{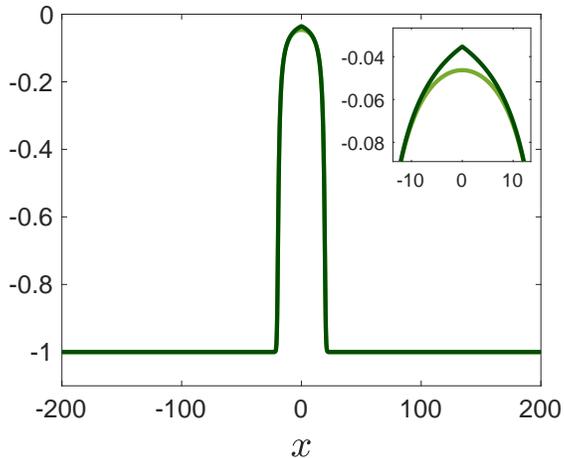}
\caption{Graph of the split-domain ansatz from Eq.~\eqref{eq:phi8_split} (dark green) and its minimized counterpart (light green).}
\label{fig:splitMinVsNon}
\end{figure}

In this case, the ansatz is continuous at $x=0$ but its first derivative is not. The minimized version for the split-domain ansatz is quite similar to the non-minimized version; the ``point'' created where the kink and antikink meet at $x=0$ (due to the discontinuity in the derivative) is smoothed by the minimization procedure, but the two look rather similar otherwise. Figure~\ref{fig:splitMinVsNon} shows how the minimized version of the split-domain ansatz differs from its non-minimized counterpart. The dynamics of the split-domain ansatz comes closest to the minimized case in that it generically leads to attraction of the kink and antikink.

One other property of the various ans{\"a}tze that is worth mentioning is the relative smoothness of the velocity graphs. One sees significant oscillations in the velocity of the center of the kink (or antikink) for all of the non-minimized ans{\"a}tze (see, e.g.,  Fig.~\ref{fig:phi8sumEvolution}). The minimized versions of all three ans{\"a}tze, on the other hand, show a steadily increasing velocity function (see, e.g.,  Fig.~\ref{fig:phi8sumMinimizedEvolution}).

\begin{figure}[tbp]
\begin{center}
\subfigure[]{
\includegraphics[scale=0.4]{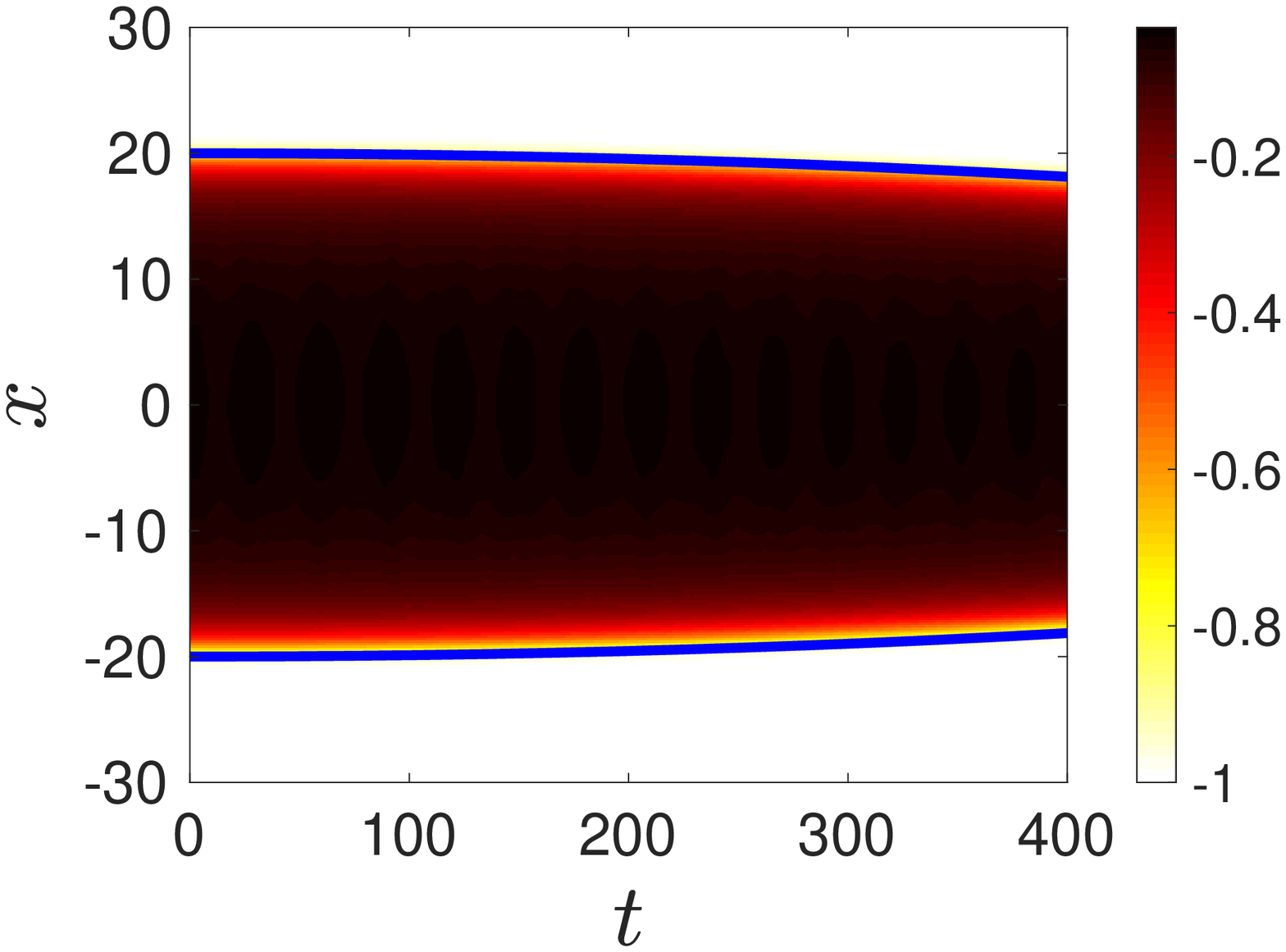}} 
\subfigure[]{
\includegraphics[scale=0.4]{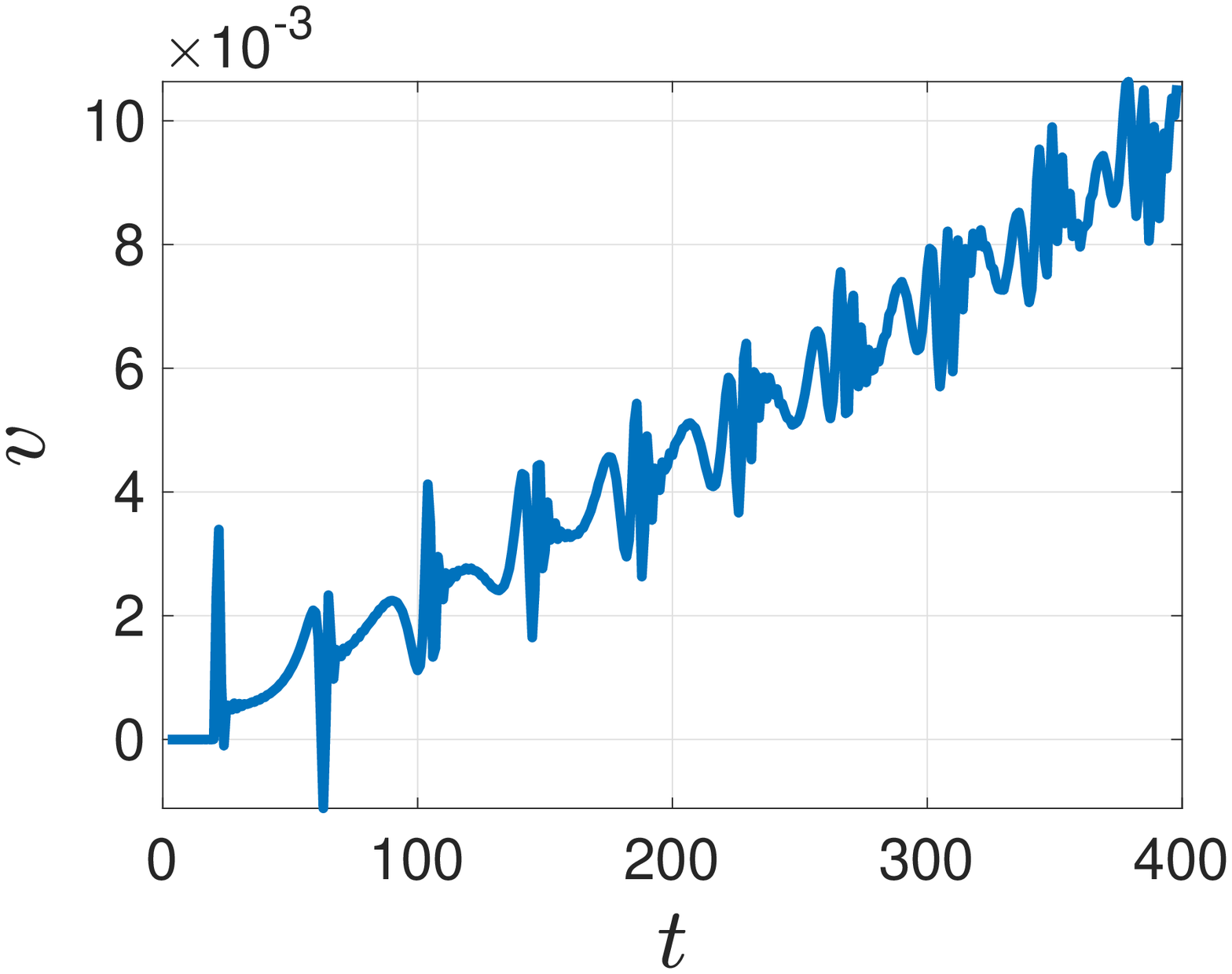}}
\subfigure[]{
\includegraphics[scale=0.4]{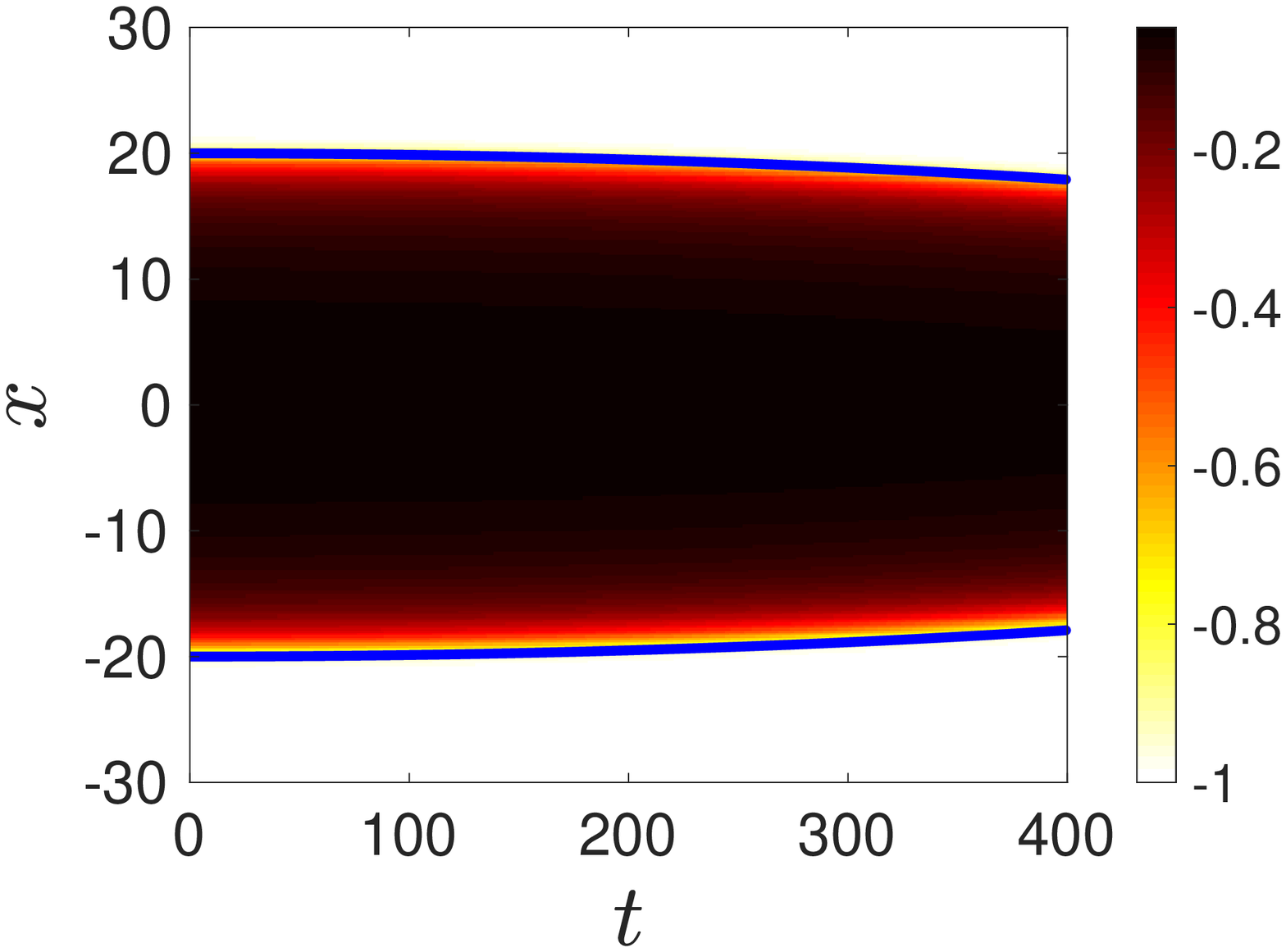}} 
\subfigure[]{
\includegraphics[scale=0.4]{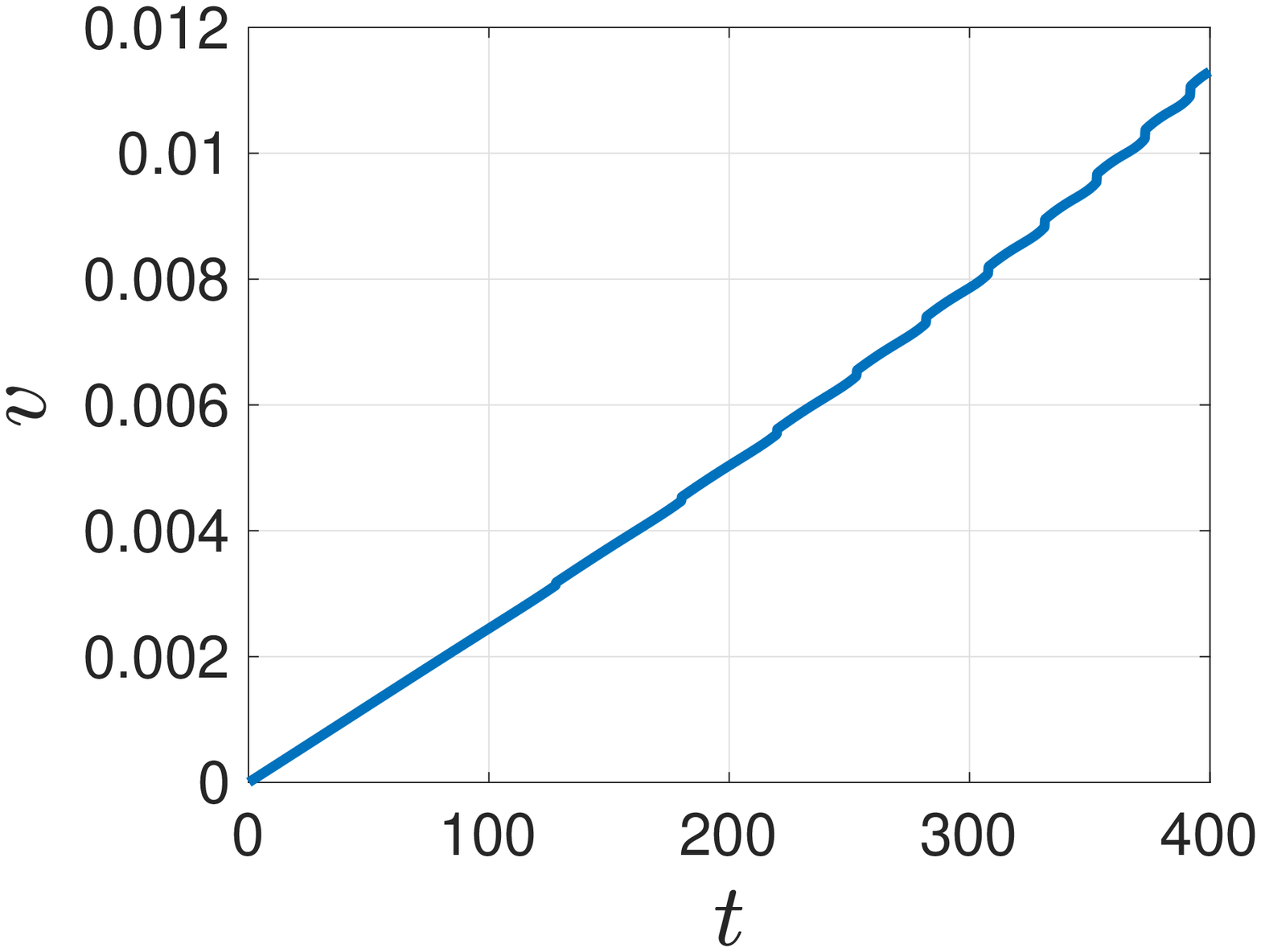}}
\end{center}
\caption{Using the split-domain ansatz from Eq.~\eqref{eq:phi8_split} (not minimized) to create initial conditions for the $\varphi^8$ model, we obtain panels (a) and (b), while using minimization of the split-domain ansatz, we obtain panels (c) and (d); panels (a) and (c) are contour plots of the solution, while panels (b) and (d) are the velocity plots stemming from solving the PDE. All panels are for $x_0=20$ and $v=0$.}
\label{fig:splitNonMinimized}
\end{figure}

In Table \ref{table:absPdePhi8},
we show a comparison of the minimized versus non-minimized values of the PDE residual, $maxAbsPde$, for the product and split-domain ans\"{a}tze. Note that for the non-minimized split domain case, $\partial^2\varphi/\partial x^2$ is not defined due to the discontinuity in the derivative at $x=0$, and hence the value of $maxAbsPde$ is listed as NA (``not available''). Another reason as to why we have elected not to provide this value is that the ansatz was constructed from the (numerically evaluated) exact solution of the BPS equation (no minimization) for each $x>0$ and $x<0$, which already satisfy $maxAbsPde=0$ numerically.  Figure~\ref{fig:splitNonMinimized} shows the equivalent plots of those in   Fig.~\ref{fig:phi8sumEvolution} (non-minimized sum ansatz) and Fig.~\ref{fig:phi8sumMinimizedEvolution} (minimized sum ansatz) for the split-domain ansatz from Eq.~\eqref{eq:phi8_split}.

\begin{table}
\begin{center}
\begin{tabular}{ c@{\hskip 12pt} c@{\hskip 12pt} c@{\hskip 12pt} c@{\hskip 12pt} c }
\hline
\hline
$x_0$ & prod (non) & prod (min) & split (non) & split (min)\\
\hline
100 & 0.0024 & $1.6\cdot 10^{-8}$ & NA & $1.5\cdot 10^{-8}$\\
50 & 0.0048 & $2.2\cdot 10^{-7}$ & NA & $2.2\cdot 10^{-7}$\\
20 & 0.012 & $8.6\cdot 10^{-6}$ & NA & $8.6\cdot 10^{-6}$\\
10 & 0.025 & $1.4\cdot 10^{-4}$ & NA & $1.4\cdot 10^{-4}$\\
5 & 0.053 & $2.0\cdot 10^{-3}$ & NA & $2.0\cdot 10^{-3}$\\
\hline
\hline
\end{tabular}
\end{center}
\caption{$maxAbsPde$ for minimized (``min'') and non-minimized (``non'') product (``prod'') and split-domain (``split'') ans\"{a}tze for the example $\varphi^{8}$ model.}
\label{table:absPdePhi8}
\end{table}

\begin{figure}[tbp]
\centering
\includegraphics[width=.5\textwidth]{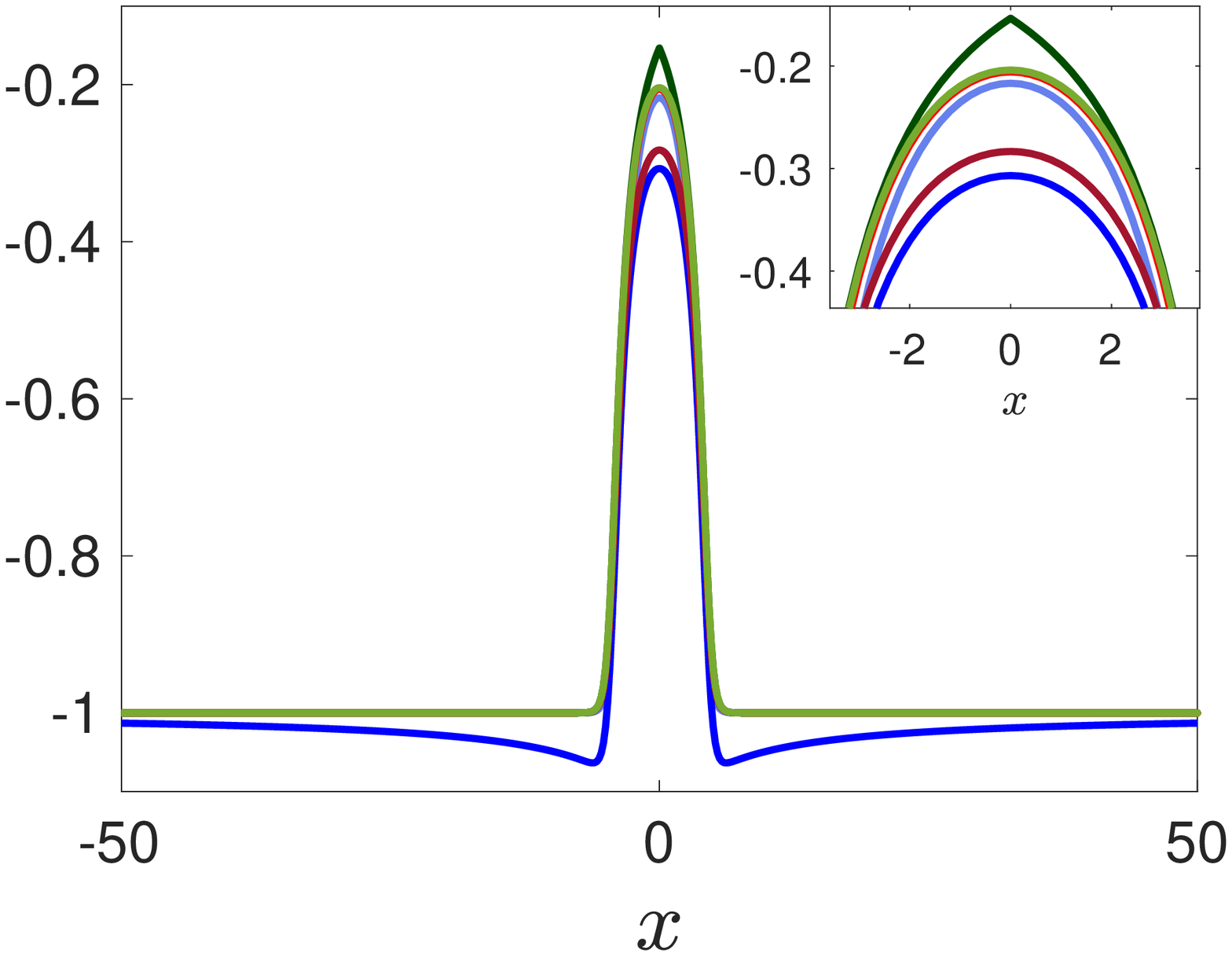}
\caption{Comparison graph of all the different kink-antikink ans{\"a}tze for the example $\varphi^8$ model and $x_0=4.5$. From top to bottom respectively: Split-domain (dark green), minimized split-domain (light green), minimized product (light red), minimized sum (light blue), product (dark red), sum (dark blue).}
\label{all_ansatzes}
\end{figure}

A plot giving a sense of how the initial conditions for the $\varphi^8$ model compare for the different ans{\"a}tze is shown in Fig.~\ref{all_ansatzes} for $x_0=4.5$ and $v = 0$. We observe that all minimized ans{\"a}tze lie between the sum/product and the split-domain ansatz. Interestingly, the results of the different minimization procedures are close to each other functionally, and the differences between them are difficult to detect without zooming in. 

To summarize: our results indicate that the correct interpretation of the nature of the pairwise kink-antikink interaction is that the kink and antikink \emph{attract} each other. We proposed the minimization procedure in Sec.~\ref{sec:improved_ic} as a way to ``distill'' the initial data and, thus, observe the genuine interaction dynamics of the kink and antikink without the detrimental side effects of the undershoot caused by their tails, as well as the radiation caused by the inexact initial conditions. While it is impossible to push the objective functional $\mathcal{I}$ from Eq.~\eqref{eq:min_funct} to zero exactly (due to the absence of multi-soliton solutions for such a non-integrable model), the minimization of $\mathcal{I}$ brings the initial $\varphi$ field as close as possible to a distilled configuration involving the superposition of a kink and an antikink. On the other hand, in the absence of access to such a minimization procedure, our recommendation is to use the split-domain ansatz from Eq.~\eqref{eq:phi8_split} directly, as it is the one that bears the least spurious byproducts among the ``standard'' multi-soliton ans{\"a}tze, even though it introduces a derivative discontinuity at $x=0$.

\subsection{Other examples: $\varphi^{10}$ and $\varphi^{12}$ models}

We find similar behaviors when considering ans{\"a}tze for the corresponding $\varphi^{10}$ and $\varphi^{12}$ field theories with three degenerate minima. In particular, we considered  the models represented by the potentials $V(\varphi) = \varphi^6 (1-\varphi^2)^2$ \cite[Sec.~IV D.3]{khare} and $V(\varphi) = \varphi^8 (1-\varphi^2)^2$ \cite[Sec.~IV D.1]{khare}, respectively. These examples come from the systematic classification of higher-order field theory potentials with degenerate minima \cite{khare} for which exact (albeit implicit) kink solutions are possible. {Using the methodology introduced in Sec.~\ref{sec:Asymptotics}, it can be shown that these potentials satisfy the conditions for the existence of kinks with power-law tails.} More specifically, a kink of the model $\varphi^{10}$ potential above approaches $-1$ exponentially as $x\rightarrow -\infty$, but approaches $0$ as $k/x^{1/2}$ (for some $k$ constant) when $x\rightarrow+\infty$. Similarly, a kink of the model $\varphi^{12}$ potential above approaches $-1$ exponentially as $x\rightarrow -\infty$, but approaches $0$ as $k/x^{1/3}$ (for some $k$ constant) when $x\rightarrow+\infty$. Thus, these higher-order field theory models possess solutions with ``fatter'' tails.

Table \ref{table:absPdePhi10} shows the $maxAbsPde$ residuals for the $\varphi^{10}$ model in a way that parallel Tables \ref{table:sum} and \ref{table:absPdePhi8} 
for the $\varphi^8$ case. 
Table \ref{table:absPdePhi12} shows the equivalent results for the $\varphi^{12}$ model. We observe similar trends for these models to what we saw in the $\varphi^8$ case; once again, the minimization procedure significantly improves the quantitative agreement between an initial condition ansatz profile and a hypothetical one that exactly satisfies the PDE~\eqref{eq:nkg}.

\begin{table}
\begin{center}
\begin{tabular}{ c@{\hskip 12pt} c@{\hskip 12pt} c@{\hskip 12pt} c@{\hskip 12pt} c@{\hskip 12pt} c@{\hskip 12pt} c }
\hline
\hline
$x_0$ & sum (non) & sum (min) & prod (non) & prod (min) & split (non) & split (min)\\
\hline
100 & 0.51 & $2.8\cdot 10^{-7}$ & 0.020 & $2.9\cdot 10^{-7}$ & NA & $2.9\cdot 10^{-7}$\\
50 & 0.85 & $2.4\cdot 10^{-6}$ & 0.028 & $2.4\cdot 10^{-6}$ & NA & $2.4\cdot 10^{-6}$\\
20 & 1.8 & $4.4\cdot 10^{-5}$ & 0.046 & $4.1\cdot 10^{-5}$ & NA & $4.1\cdot 10^{-5}$\\
10 & 3.5 & $4.4\cdot 10^{-4}$ & 0.067 & $3.8\cdot 10^{-4}$ & NA & $3.7\cdot 10^{-4}$\\
5 & 7.4 & $6.1\cdot 10^{-3}$ & 0.10 & $3.7\cdot 10^{-3}$ & NA & $3.4\cdot 10^{-3}$\\
\hline
\hline
\end{tabular}
\end{center}
\caption{$maxAbsPde$ for minimized (``min'') and non-minimized (``non'') sum, product (``prod'') and split-domain (``split'') ans\"{a}tze applied to the example $\varphi ^{10}$ model.}
\label{table:absPdePhi10}
\end{table}

\begin{table}
\begin{center}
\begin{tabular}{ c@{\hskip 12pt} c@{\hskip 12pt} c@{\hskip 12pt} c@{\hskip 12pt} c@{\hskip 12pt} c@{\hskip 12pt} c }
\hline
\hline
$x_0$ & sum (non) & sum (min) & prod (non) & prod (min) & split (non) & split (min)\\
\hline
100 & 3.0 & $5.6\cdot 10^{-6}$ & 0.039 & $1.6\cdot 10^{-6}$ & NA & $7.6\cdot 10^{-7}$\\
50 & 5.1 & $5.3\cdot 10^{-6}$ & 0.049 & $5.1\cdot 10^{-6}$ & NA & $5.1\cdot 10^{-6}$\\
20 & 11.4 & $7.7\cdot 10^{-5}$ & 0.067 & $6.8\cdot 10^{-5}$ & NA & $6.7\cdot 10^{-5}$\\
10 & 22.5 & $7.5\cdot 10^{-4}$ & 0.087 & $5.2\cdot 10^{-4}$ & NA & $5.0\cdot 10^{-4}$\\
5 & 48.5 & $1.7\cdot 10^{-2}$ & 0.116 & $4.4\cdot 10^{-3}$ & NA & $3.9\cdot 10^{-3}$\\
\hline
\hline
\end{tabular}
\end{center}
\caption{$maxAbsPde$ for minimized (``min'') and non-minimized (``non'') sum, product (``prod'') and split-domain (``split'') ans\"{a}tze applied to the example $\varphi ^{12}$ model.}
\label{table:absPdePhi12}
\end{table}

Though the contour and velocity plots for the $\varphi^{10}$ and $\varphi^{12}$ models are generally quite similar to the $\varphi^8$ plots for many cases, we point out a few cases in which the collisions in the higher-order field theories differ. Primarily, the differences occur for the non-minimized sum ansatz, for which we find that the initial conditions chosen based on this ansatz do not lead to clearly attracting or repelling kink-antikink pairs, for certain values of $x_0$. Rather, for these cases, the ansatz leads to solutions that show oscillations in the range $\varphi=-1$ to $\varphi=1$ rather than topological solitons connecting $\varphi=-1$ to $\varphi=0$. 

A way to explain this strange result is to consider the fact that for such fatter-tail cases, as the ones arising from the example $\varphi^{10}$ and $\varphi^{12}$ field theories considered herein, the undershoot of the kinks is so substantial that not only is the $\varphi=0$ fixed point not reached between the kinks, but also  neither is the asymptotic value of $\varphi=-1$ as $|x| \gg 1$. In other words, the (non-minimized) sum ansatz provides an extremely poor initial conditions for the fatter-tail cases.

\begin{figure}[tbp]
\centering
\includegraphics[width=0.5\textwidth]{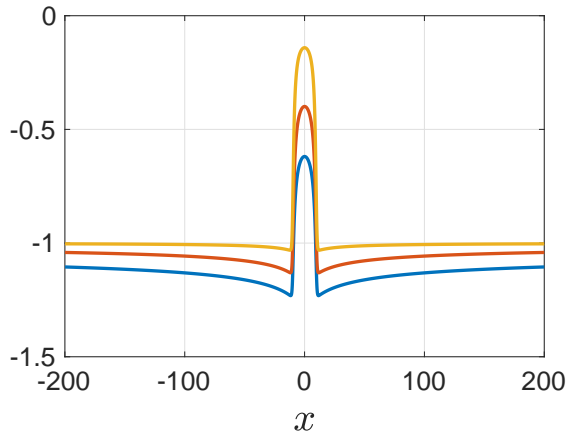}
\caption{Graphs of initial conditions generated for the $\varphi^8$ (top curve), $\varphi^{10}$ (middle curve) and $\varphi^{12}$ (bottom curve) models using the sum ansatz (no minimization), showing the worsening quality of representation of the kink-antikink configuration.}
\label{fig:initialsX0-10}
\end{figure}

We can obtain some further insight into this quantitative disagreement by considering the graphs of the sum ansatz for the three cases; see Fig.~\ref{fig:initialsX0-10} ($\varphi^8$ on top, $\varphi^{10}$ in middle, and $\varphi^{12}$ at the bottom). The ``fat tails'' of the $\varphi^{12}$ kink give such a large boost to the points in the middle (springboard effect) that they travel past the potential minimum at $\varphi=0$ to the neighborhood of $\varphi=1$ under the evolution of the PDE~\eqref{eq:nkg}. This effect persists  for $x_0=5$, $10$, $20$, $50$ and $100$ for the $\varphi^{12}$ model (recall Table~\ref{table:absPdePhi12}), and to a lesser extent for $x_0=5$, $10$ and $20$ for the $\varphi^{10}$ model (recall Table~\ref{table:absPdePhi10}). In all of these cases, the notions of ``attracting kinks'' and ``repelling kinks'' are no longer meaningful.

\section{Potential Energy and Force of Interaction as a Function of Separation Distance}
\label{sec:force}

\begin{figure}[tbp]
\centering
\includegraphics[width=0.5\textwidth]{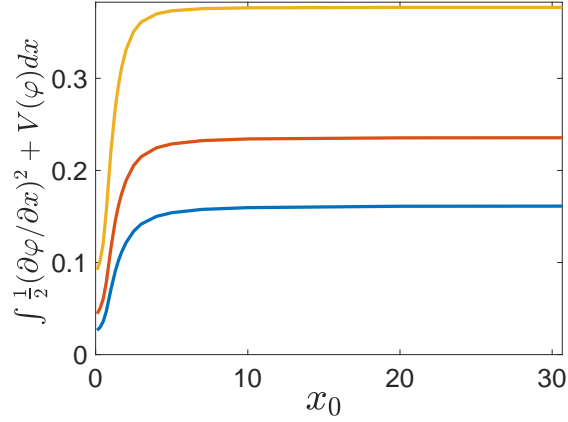}
\caption{Graphs of the field's potential energy $\int \frac{1}{2}(\partial\varphi/\partial x)^{2}+V(\varphi)dx$ of an initial condition $\varphi$ as a function of $x_0$ for the $\varphi^8$ (top curve), $\varphi^{10}$ (middle curve) and $\varphi^{12}$ (bottom curve) models, all calculated using the minimized split-domain ansatz.}
\label{fig:landscape}
\end{figure}

Lastly, we can illustrate the attractive nature of the force between a kink and antikink in two additional ways. In Fig.~\ref{fig:landscape}, we show the potential energy  $E[\varphi]$ as defined in Eq.~\eqref{eq:energ} (assuming a stationary solution as before, so $\partial\varphi/\partial t =0$), for a kink-antikink pair; i.e., we plot $\int \frac{1}{2}(\partial\varphi/\partial x)^{2}+V(\varphi)\,dx$ as a function of $x_0$. Here, we employ only the minimized split-domain ansatz, which we concluded above was the most accurate.
 
\begin{figure}[tbp]
\centering
\includegraphics[width=0.5\textwidth]{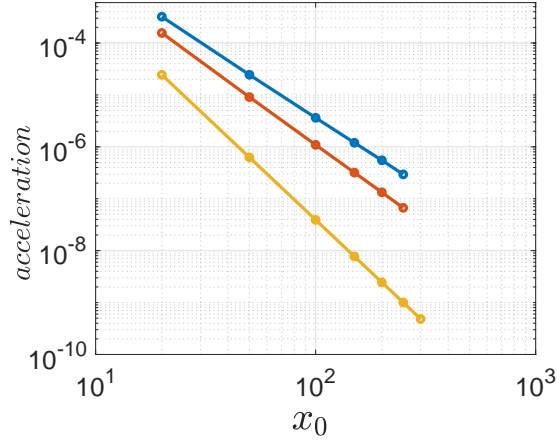}
\caption{Log-log plots of the kink's acceleration computed from the PDE evolution simulation data, and the corresponding fitted power-law model, as a function of $x_0$, for the $\varphi^8$ (bottom curve), $\varphi^{10}$ (middle curve) and $\varphi^{12}$ (top curve) models. All simulations used initial conditions generated using the minimized split-domain ansatz.}
\label{fig:powerLaw}
\end{figure}

Also, we have calculated the acceleration of the left kink (a proxy for the kink-antikink force of interaction), from the previously computed kink velocity $v$ from the PDE evolution, as function of $x_0$ for a minimized split-domain initial condition ansatz and zero initial velocity. In this case, the acceleration is quite steady for a short period of time. Six data points were collected for the $\varphi^{10}$ and $\varphi^{12}$ models, and seven for the $\varphi^8$ model, with $x_0\in[ 20, 300]$, and a power-law model $ax_0^{-b}$ was fit to the data. In all cases, an excellent fit was obtained. Specifically, we find $b=3.998 \pm 0.002$ for the chosen $\varphi^8$ model,  $b=3.067 \pm  0.019$ for the chosen $\varphi^{10}$ model, and  $b=2.764 \pm 0.025$ for the chosen $\varphi^{12}$ model, all within the 95\% confidence interval for the fit. Figure \ref{fig:powerLaw} shows the simulation data and fits on a log-log plot, for all three cases.
  
We note that this scaling (i.e., $b\approx4$ for the example $\varphi^8$ model) of the acceleration (thus, the force of interaction) with the half-separation is in line with the theoretical prediction for the $-4$ power-law decay of the force for the same $\varphi^8$ model in~\cite{gani17} and, more recently, in {\cite{mantonnow,mantonnow2}}.  A systematic study of this interaction force and its dependence on the kink-antikink separation, for arbitrary power-law tails, is the subject of future work {\cite{our_force_preprint}}, as it is a topic of interest in its own right; we would digress from the main theme of the present study if we were to pursue it here.

\section{Conclusions and Future Work}
\label{sec:conclusions}

In the present work, we have systematically interrogated the dynamics of kink-antikink interactions in higher-order polynomial field theoretic models (of eight degree and higher) with degenerate minima. The specific feature of these models that we have sought to capture is the presence of \emph{long-range interactions} via kink tail asymptotics that do \emph{not} decay exponentially but rather decay algebraically. Although a $\varphi^8$ model was used as our featured example, we also demonstrated that our discussion of how to properly generate initial conditions for direct numerical simulations of kink-antikink interactions applies to $\varphi^{10}$ and $\varphi^{12}$ models with ``fatter tails.'' Our main finding was that, for all of these higher-order field theories, the standard sum ans{\"a}tze (of a kink plus an antikink, spaced some distance apart) for direct numerical simulation of collision are problematic. These ans{\"a}tze exhibit a significant undershoot around the two kinks in the combined profile, which leads to considerable radiation in the numerical solution for $t>0$. These unwanted effects, in turn, are responsible for the apparent observation of unwarranted features such as repulsive dynamics (for a sum ansatz) or transitions between repulsive and attractive dynamics (for a product ansatz). We have argued that, among the simpler ans{\"a}tze available, the one leading to the most realistic kink-antikink interaction dynamics (i.e., the final results are not contaminated by the details of the initial conditions for a simulation) is the one we have termed the \emph{split-domain ansatz}. Moreover, we have argued towards the usefulness of a suitable minimization procedure that ``distills'' a given ansatz further by making a closer match to a stationary kink-antikink solution of the problem. The minimization procedure reduces the undershoots in the combined profile and, thus, reduces  radiation wavepackets (as well as their side effects) in simulations. Once an initial condition was thus suitably prepared, we observed attraction between a kink and an antikink in all of the higher-order field theories (i.e., our prototypical $\varphi^8$ example, as well as $\varphi^{10}$ and $\varphi^{12}$ models), much like in the classical $\varphi^4$ field theory. This type of improvement in the kink-antikink state construction allowed us also to unambiguously obtain the power law nature of the kink-antikink interaction force and how its exponent varies among the different higher-order field theories examines.

Naturally, this study opens up numerous avenues for future work on the interactions of topological solitons in higher-order field theories. The most canonical extension of this work concerns the outcome of collisional events for different initial speeds of the kink and antikink (in this work, we took $v=0$ in all of our examples), and across our proposed variety of initial condition ans{\"a}tze. Such an exploration and a corresponding systematic study will be reported elsewhere in the future. Another open question is: how much of the above-described interaction picture can be captured through a semi-analytical approximation such as the method of collective coordinates (CC)? A first attempt is given in the Appendix that follows, yet as can be seen there, it is somewhat limited in its ability to capture in detail the kink-antikink interactions. Moreover, at the present stage, the CC model is lacking the inclusion of the internal vibration mode of the kink; incorporating the latter appears to be extremely cumbersome in the present setup. Lastly, another important question is: how much of the above-described phenomenology can be captured in an experiment? We are not immediately aware of experiments involving higher-order field theories. However, for complex variants of the $\varphi^4$ field theory, such as nonlinear Schr{\"o}dinger models, kinks can be introduced via interference events~\cite{weller} or imprinting processes~\cite{becker}, among others. In all of these examples, creation of kinks is accompanied by radiation and by tails. It is then natural to ask: to what extent can long-range interactions of kinks and antikinks be captured in a realistic experimental setup? The answer to such questions is currently under consideration and will be reported in future publications.

\section*{Acknowledgments}

I.C.C.\ and P.G.K. thank A.\ Khare and A.\ Saxena for illuminating discussions on higher-order field theories. I.C.C.\ also thanks the Purdue Research Foundation for an International Travel Grant, which allowed him to visit V.A.G.\ and R.V.R.\ in Moscow and continue this work. The work of V.A.G.\ was supported by the MEPhI Academic Excellence Project under contract No.~02.a03.21.0005, 27.08.2013. This material is based upon work supported by the  National Science Foundation under Grant No.\ PHY-1602994 and under Grant No.\ DMS-1809074 (P.G.K.).

\appendix

\section*{Appendix: Collective Coordinate Approach and Connection to the Numerical Results}

In this Appendix, we apply the method of collective coordinates (CC) using the minimized initial condition ans{\"a}tze, which we introduced in the main text above to analyze the kink-antikink interactions numerically. To this end, recall that the Lagrangian for our neutral scalar field theories is
\begin{eqnarray}\label{eq:lagr_int}
L = \int_{-\infty}^{+\infty} \mathscr{L}\,dx = \int_{-\infty}^{+\infty} \left[\frac{1}{2} \left( \frac{\partial\varphi}{\partial t} \right)^2-\frac{1}{2} \left( \frac{\partial\varphi}{\partial x} \right) ^2-V(\varphi)\right] dx,
\end{eqnarray}
where the Lagrangian density is given in Eq.~\eqref{eq:largang} and the potential is given in Eq.~\eqref{eq:V_8} for the chosen $\varphi^8$ model. Now, for all minimized ans{\"a}tze, we can reduce the PDE~\eqref{eq:nkg} to a Hamiltonian dynamical system with one degree of freedom as follows. First, we obtain an \emph{effective} Lagrangian by evaluating Eq.~\eqref{eq:lagr_int} using the ansatz for $\varphi$, having identified ``$x-vt$'' as the collective coordinate $X(t)$. The manipulation is formally denoted as
\begin{align}\label{lag1}
L_\mathrm{eff} = b_0(X)\dot X^2 - b_1(X),
\end{align}
where different ans{\"a}tze yield different functions $b_0(X)$ and $b_1(X)$. In the following subsections, we will present the formul\ae\ for these coefficients for each ansatz. The Euler--Lagrange equation rendering the functional $L_\mathrm{eff}$ in Eq.~\eqref{lag1} stationary is
\begin{equation}\label{euler-lagrange1}
\frac{\partial L_\mathrm{eff}}{\partial X} - \frac{d}{dt}\bigg(\frac{\partial L_\mathrm{eff}}{\partial \dot X}\bigg) = 0.
\end{equation}
The resulting dynamical evolution equation, written as a first-order system, is
\begin{subequations}
\begin{align}
\dot X &= Y, \\
\dot Y &= -\frac{1}{2}\frac{b'_0(X)}{b_0(X)}Y^2 - \frac{1}{2}\frac{b_1'(X)}{b_0(X)}.
\end{align}\label{eq:EL}\end{subequations}

We solve this first-order ODE system~\eqref{eq:EL}, subject to the initial conditions $X(0) = x_0$, $Y(0) = 0$ (corresponding to $v=0$). As before, $x_0$ is the initial half-separation between the kink and the antikink, and it is assumed that the initial speed of the kink and antikink is zero. For integration of the system, we use MATLAB's {\tt ode45} differential equations solver with adaptive time stepping and error control.

\subsection{CC method for the improved (minimized) sum ansatz}

\begin{figure}
\centering
\subfigure[]{
\includegraphics[width=0.45\textwidth]{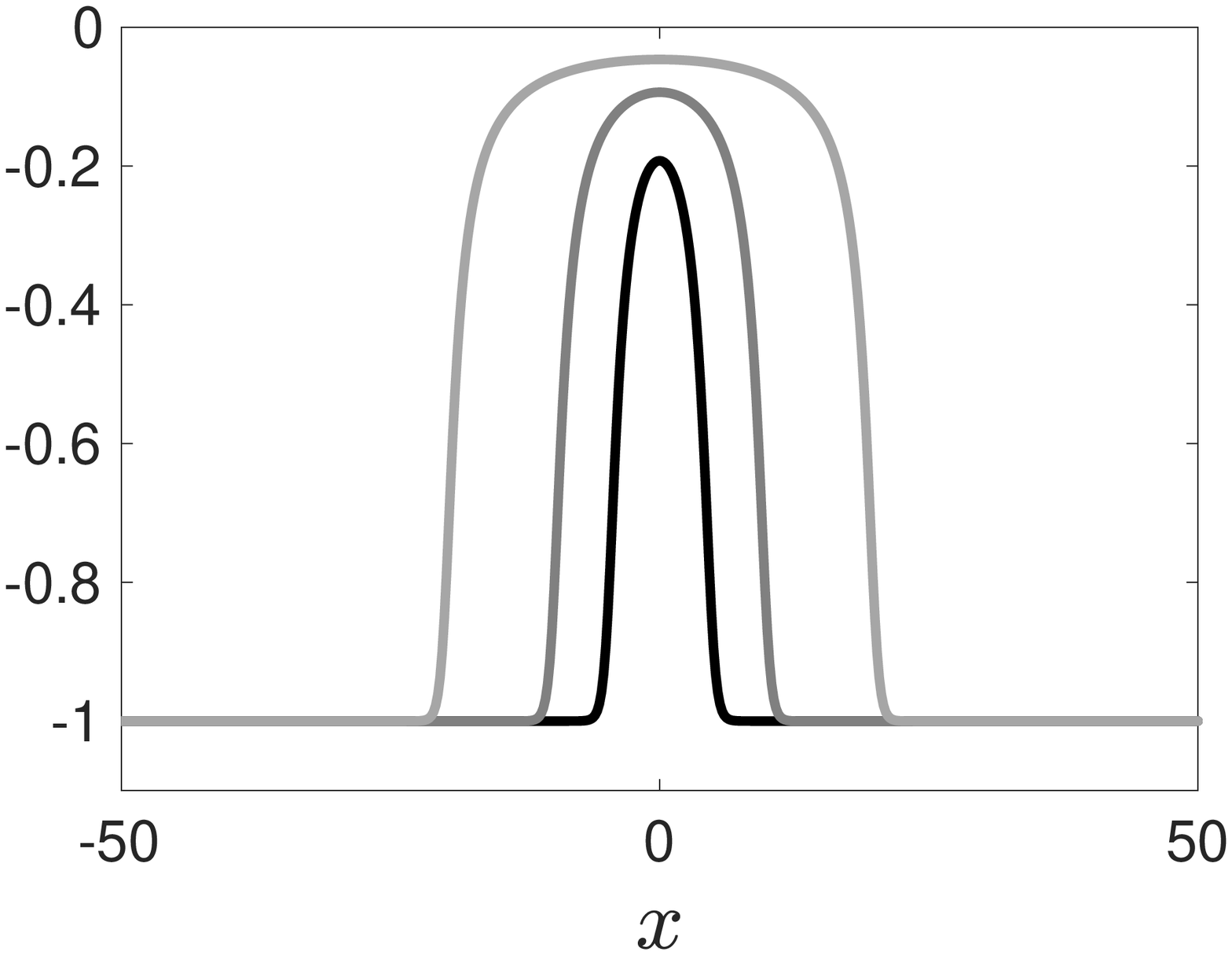}} 
\subfigure[]{
\includegraphics[width=0.45\textwidth]{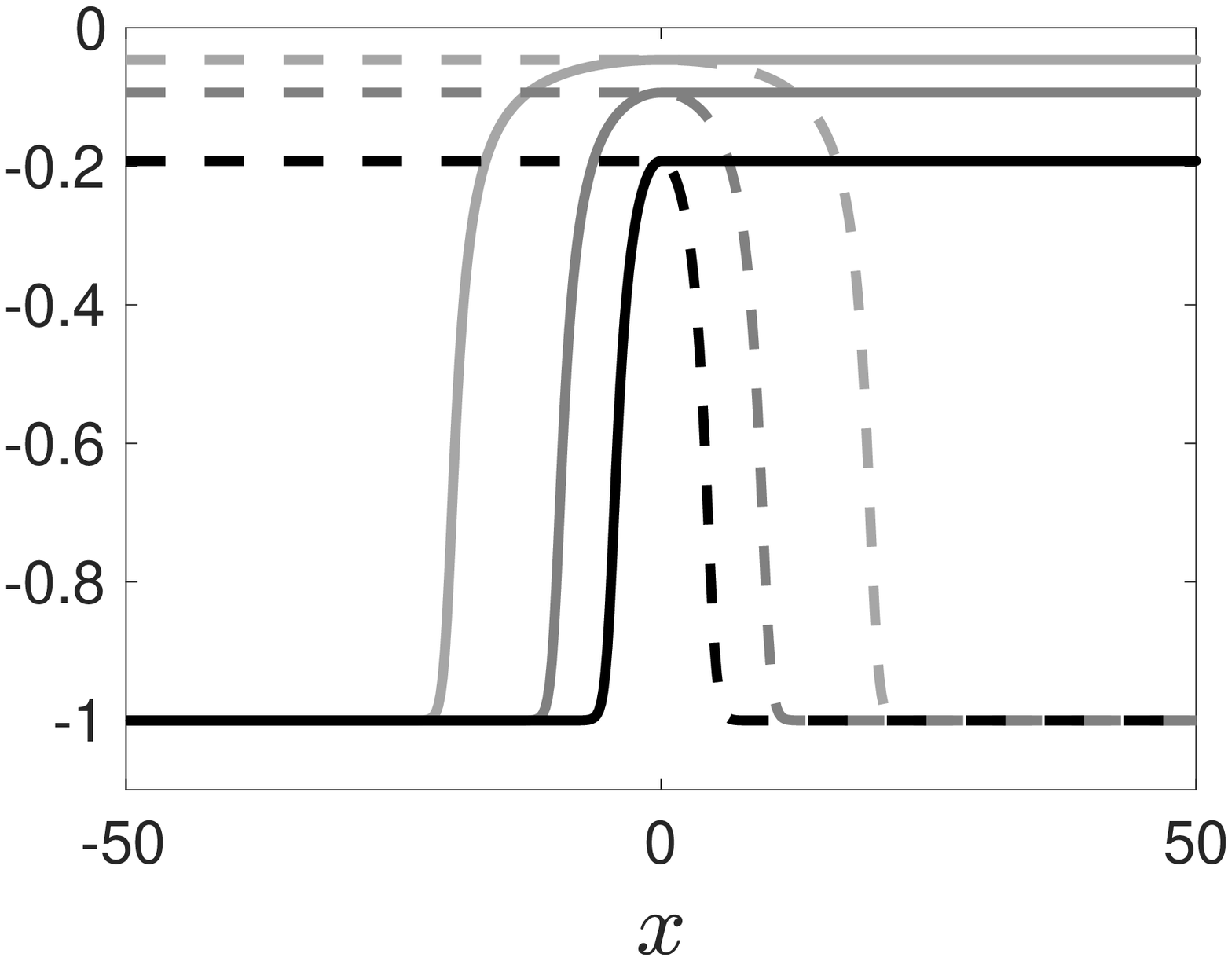}}
\caption{In (a), curves represent an initial condition, $\varphi(x,0)$, for a direct numerical simulation under the $\varphi^8$ model that has been generated via minimization of the sum ansatz, i.e., Eq.~\eqref{phi8sum_eqn}. In (b), solid curves represent $K_{a_1}(x)$, and dashed curves represent $K_{a_2}(x)$. In both panels, $x_0$ is varied from 5 to 10 to 20 (darker color curves to lighter color curves, respectively).}
\label{min-addition-cc}
\end{figure}

Let $f_a(x)$ be the function obtained by using the minimization of sum ansatz (corresponding to light blue curve in Fig.~\ref{all_ansatzes} for $x_0=4.5$) for the $\varphi^8$ model, i.e., Eq.~\eqref{phi8sum_eqn}. Then, assume a colliding kink-antikink scenario with the following field configuration
\begin{equation}\label{add-cc}
\varphi(x,t)=K_{a_1}(x+X(t)-x_0)+K_{a_2}(x-X(t)+x_0)-f_a(0),
\end{equation}
where $X(t)$ is the half-distance between the kink and antikink and
\begin{equation}
K_{a_1}(x) = \left\{
        \begin{array}{ll}
            f_a(x), & \quad x \leq 0 \\
            f_a(0), & \quad x > 0
        \end{array}
    \right., \quad\text{and}\quad
  K_{a_2}(x) = \left\{
        \begin{array}{ll}
            f_a(0), & \quad x < 0 \\
            f_a(x), & \quad x \geq 0
        \end{array}
    \right..  
\end{equation}
Observe that when $X(0)=x_0$, we have $\varphi(x,0)=f_a(x)$, which implies that the initial conditions for the $\varphi^8$ equation of motion~\eqref{eq:nkg} (i.e., ``PDE model'') and the CC approach (i.e., ``ODE model'') match. Figure \ref{min-addition-cc} presents such functions for $x_0 = 5$, $10$, $20$.

Using the ansatz in Eq.~\eqref{add-cc}, and defining $K_{a_1}^+=K_{a_1}(x+X(t)-x_0)$ and $K_{a_2}^-=K_{a_2}(x-X(t)+x_0)$, we calculate the coefficient functions in Eq.~\eqref{lag1} as follows:
\begin{align}
b_0(X) &= \frac{1}{2}\int (K_{a_1}'^+-K_{a_2}'^-)^2 \, dx,\\
b_1(X) &= \frac{1}{2}\int (K_{a_1}'^++K_{a_2}'^-)^2  \, dx +
\int V\big(K_{a_1}^++K_{a_2}^--f_a(0)\big) \, dx,
\end{align}
where the integration is to be over  $(-\infty,+\infty)$ (or a suitably large $x$-interval for numerical purposes).

\begin{figure}
\begin{center}
\subfigure[]{
\includegraphics[scale=0.4]{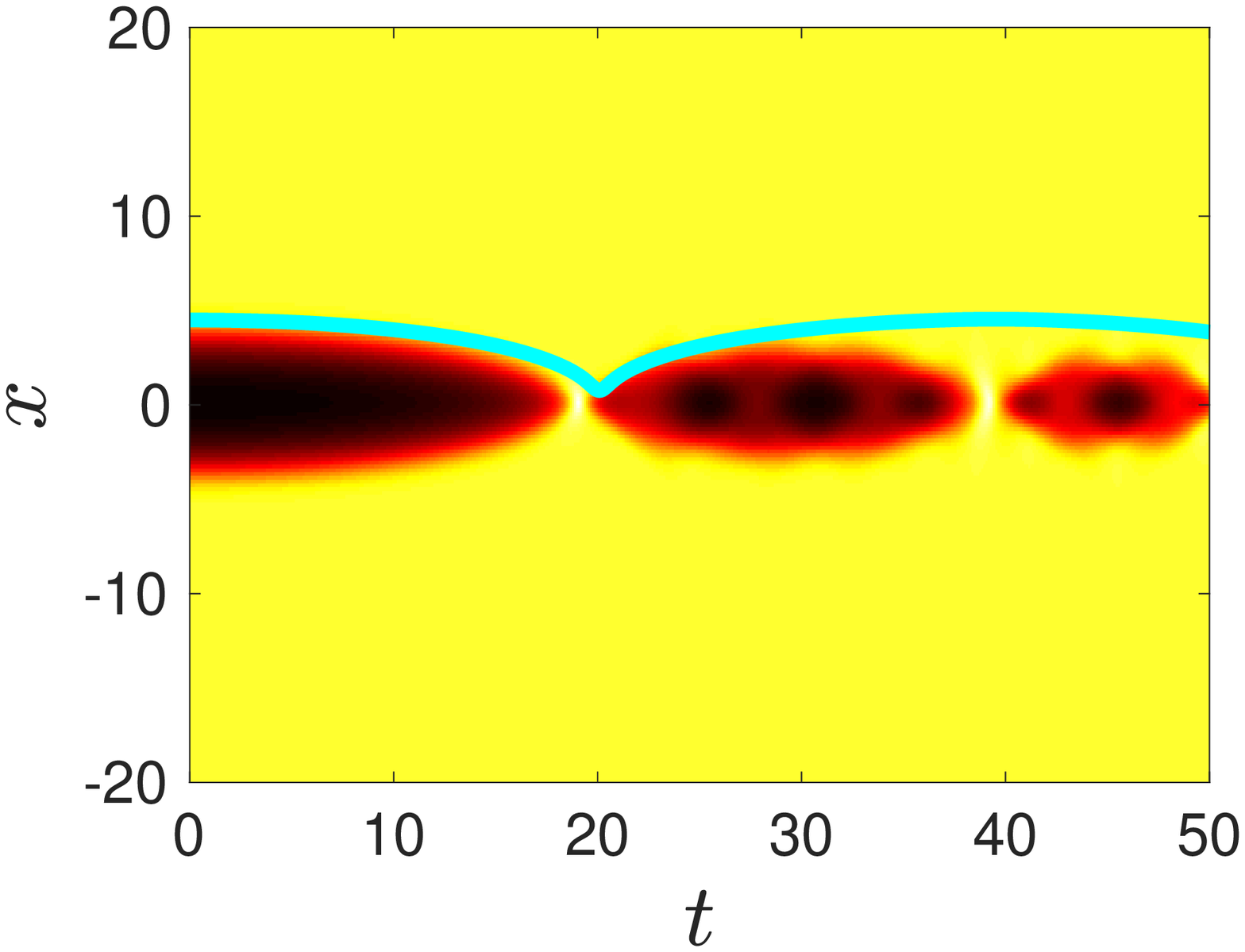}} 
\subfigure[]{
\includegraphics[scale=0.4]{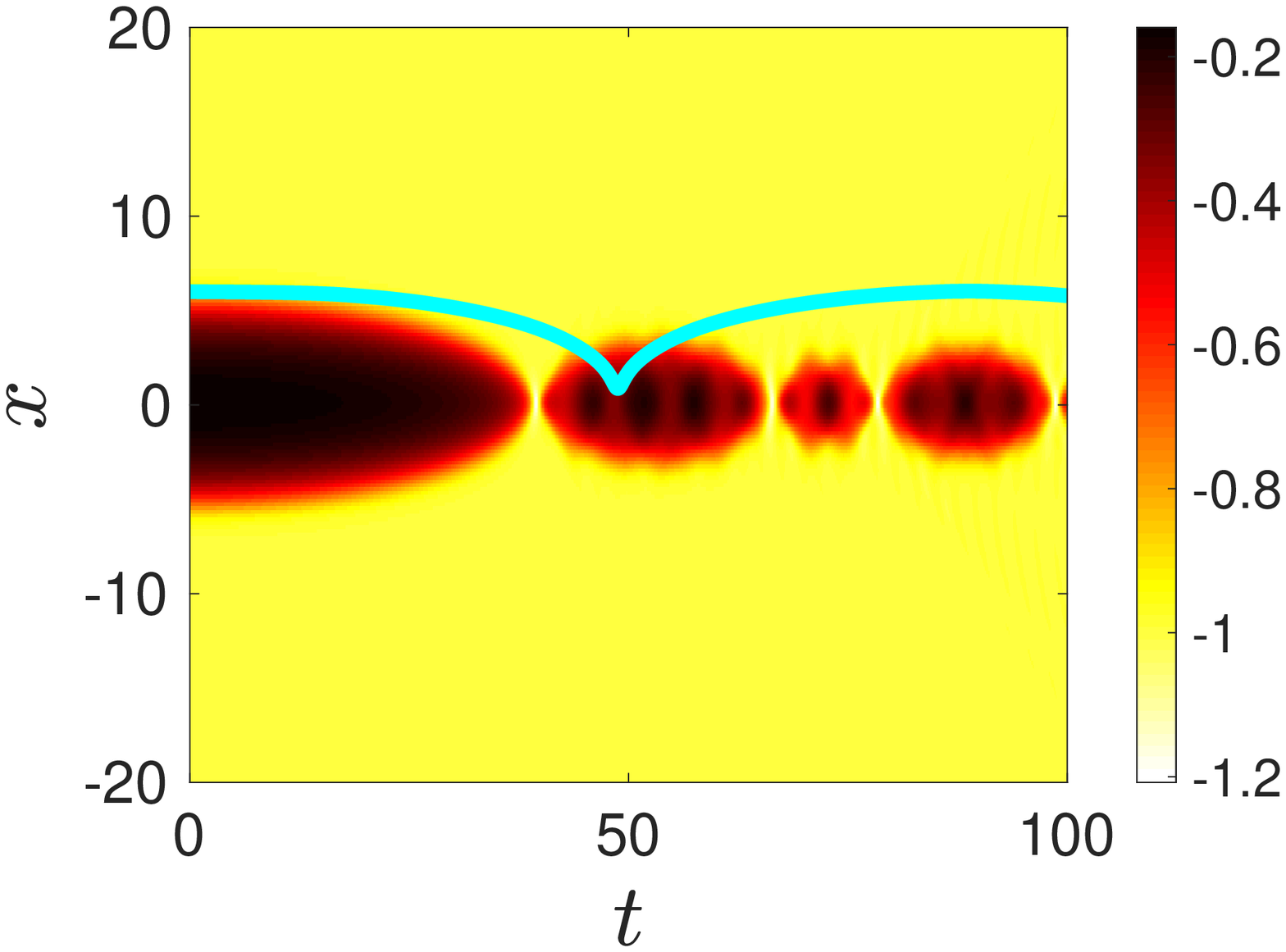}}
\end{center}
\caption{Using the minimized sum ansatz, overlays of the ODE model~\eqref{eq:EL} solution $X(t)$ (solid bold curve) on top of PDE model contour plot of $\varphi(x,t)$ for the evolution of an initially stationary ($v = 0$) kink-antikink configuration with (a) $x_0=4.5$ and (b) $x_0=6$.}
\label{addition-cc}
\end{figure}

When the initial velocity is zero, both the ODE model and the PDE model predict attraction for $x_0<x_c$, where $6<x_c<7$. The ODE model results agree better with the PDE model results as $x_0$ becomes smaller, as shwon in Fig.~\ref{addition-cc}.

\subsection{CC method for the improved (minimized) product ansatz}

As in the previous subsection, here we split the function that we obtain by using the minimization of the product ansatz for the $\varphi^8$ model. If we call that function $f_p(x)$ (corresponds to the light red curve in Fig.~\ref{all_ansatzes} for $x_0=4.5$), then we define a colliding kink-antikink system with the following field configuration
\begin{equation}\label{mul-cc}
\varphi(x,t)=\frac{1}{(f_p(0)+1)}[K_{p_1}(x+X(t)-x_0)+1)(K_{p_2}(x-X(t)+x_0)+1]-1,
\end{equation}
where $X(t)$ is the half-distance between the kink and antikink and 
\begin{equation}
K_{p_1}(x) = \left\{
        \begin{array}{ll}
            f_p(x), & \quad x \leq 0 \\
            f_p(0), & \quad x > 0
        \end{array}
    \right., \quad\text{and}\quad 
  K_{p_2}(x) = \left\{
        \begin{array}{ll}
            f_p(0), & \quad x < 0 \\
            f_p(x), & \quad x \geq 0
        \end{array}
    \right..  
\end{equation}
Observe that when $X(0)=x_0$, we have $\varphi(x,0)=f_p(x)$, which implies that the initial conditions for the $\varphi^8$ field theory's equation of motion (``PDE model'') and the CC method (``ODE model'') both match. Using the ansatz in Eq.~\eqref{mul-cc}, and defining $K_{p_1}^+=K_{p_1}(x+X(t)-x_0)$ and $K_{p_2}^-=K_{p_2}(x-X(t)+x_0)$, we calculate the coefficient functions in Eq.~\eqref{lag1} as follows:
\begin{align}
b_0(X) &= \frac{1}{2(f_p(0)+1)}\int \left[K_{p_1}'^+(K_{p_2}^-+1)-(K_{p_1}^++1) K_{p_2}'^-\right]^2 \, dx,\\
b_1(X) &= \frac{1}{(f_p(0)+1)}\left\{\frac{1}{2} \int \left[K_{p_1}'^+(K_{p_2}^-+1)+(K_{p_1}^++1) K_{p_2}'^-\right]^2 \, dx\right. \nonumber\\
&\phantom{= \frac{1}{(f_p(0)+1)}\Bigg\{} \left. + \int V\big((K_{p_1}^++1) (K_{p_2}^-+1)-1\big) \, dx\right\}.
\end{align}

\begin{figure}[tbp]
\begin{center}
\subfigure[]{
\includegraphics[scale=0.4]{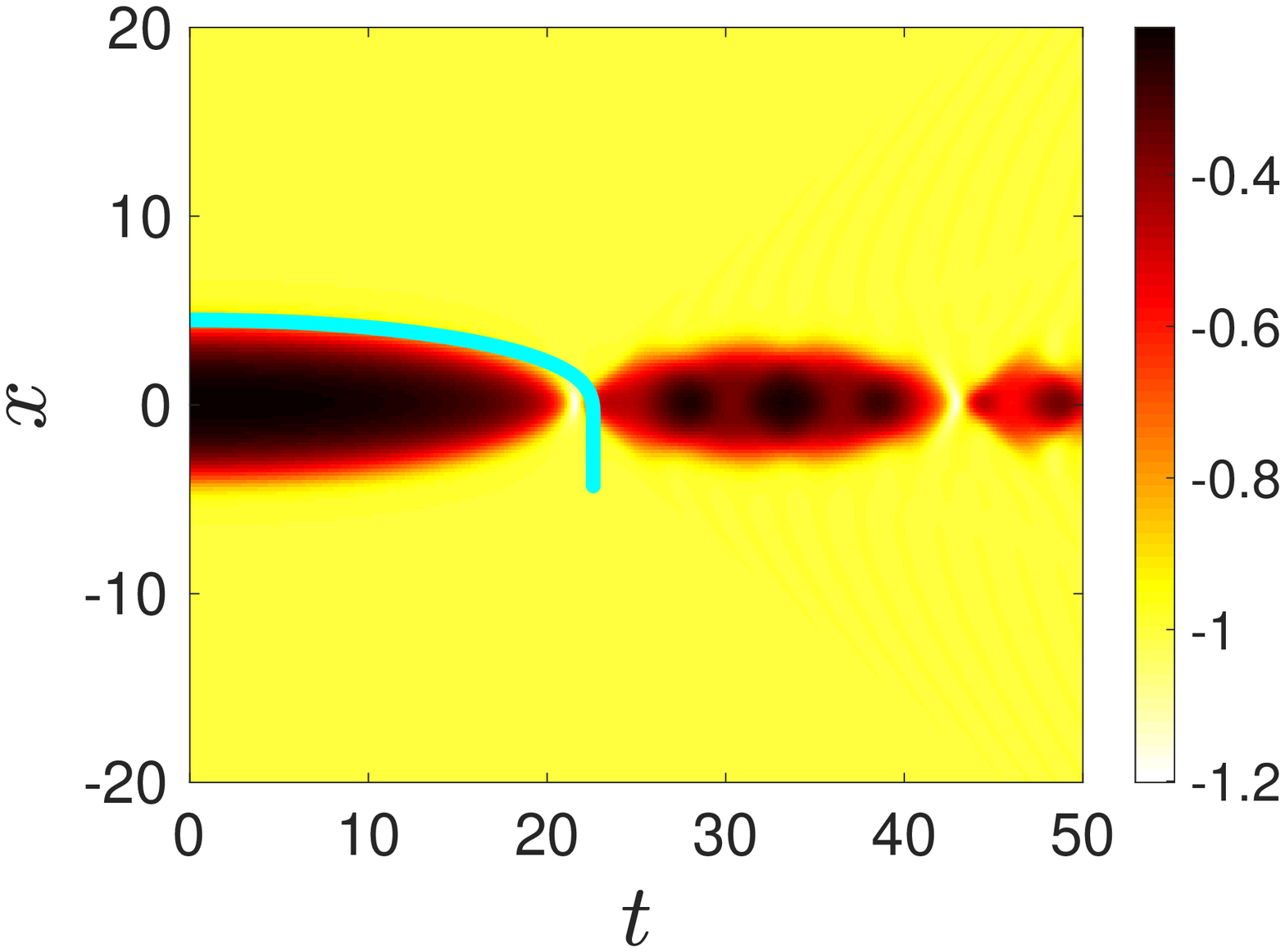}} 
\subfigure[]{
\includegraphics[scale=0.4]{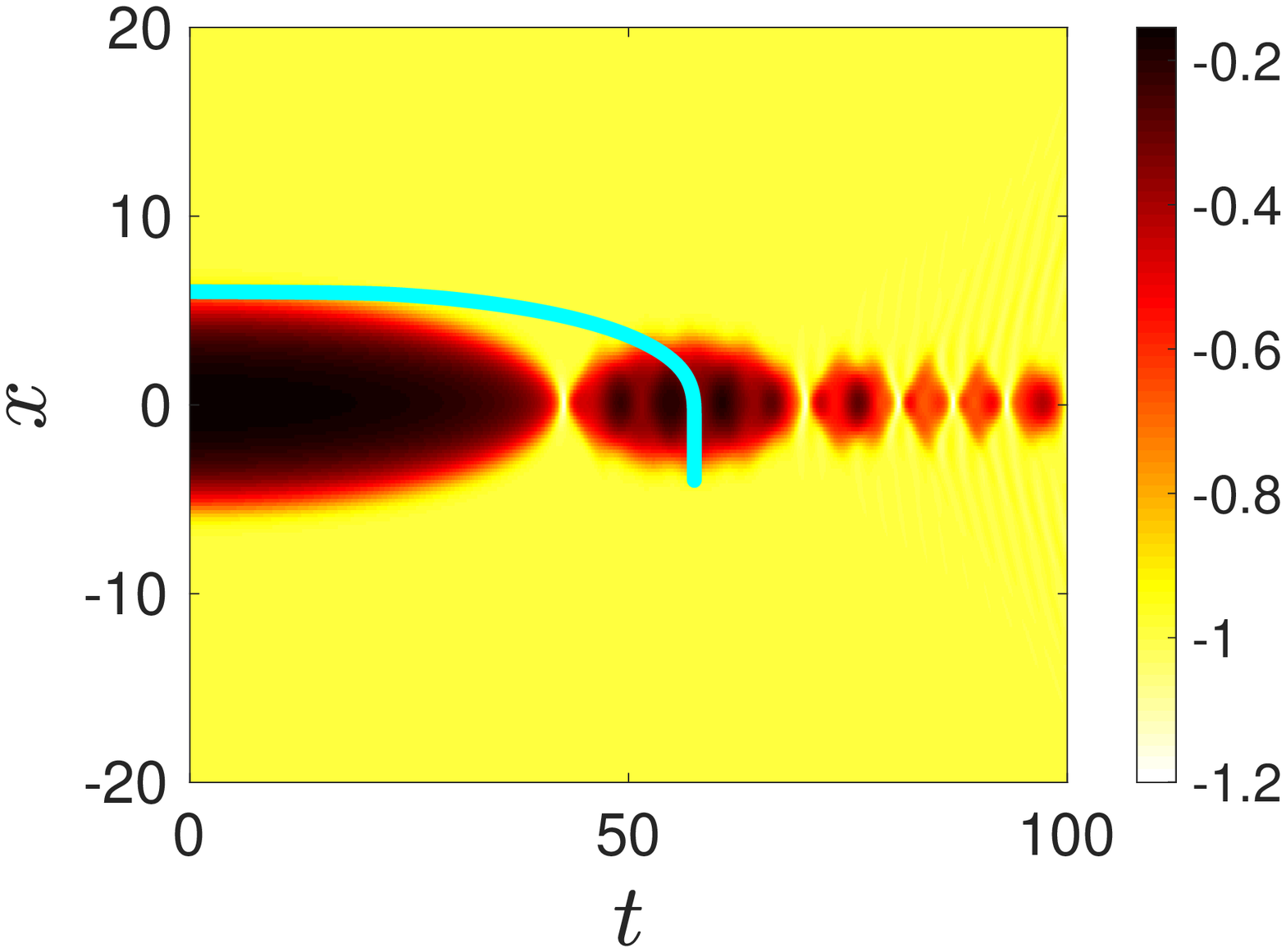}}
\end{center}
\caption{Using minimized product ansatz, overlays of the ODE model~\eqref{eq:EL} solution $X(t)$ (solid line curve) on top of PDE model contour plot for the evolution of an initially stationary ($v = 0$) kink-antikink configuration with (a) $x_0=4.5$ and (b) $x_0=6$.}
\label{multiplication-cc}
\end{figure}

For a zero initial velocity, both the ODE model and the PDE model show attraction for $x_0<x_c$, where $6<x_c<7$. The ODE model's results agree better with the PDE results as $x_0$ becomes smaller. Figure~\ref{multiplication-cc}(a) shows the ODE and PDE agreement, for $x_0=4.5$ and $v=0$, until the kink in the ODE model is expelled from the system. Meanwhile, Fig.~\ref{multiplication-cc}(b) shows attraction for $x_0=6$ and $v=0$; however, for this value of $x_0$ the agreement between the ODE and PDE mdoels is not as good.

\subsection{CC method for the improved (minimized) split-domain ansatz}

Again, we split the function that we obtain by using the minimization of split-domain ansatz for the $\varphi^8$ model. If we denote that function as $f_s(x)$ (corresponds to light green curve in Fig.~\ref{all_ansatzes} for $x_0=4.5$), then we define  a colliding kink-antikink system with the following field configuration:
\begin{equation}\label{sp-cc}
\varphi(x,t) = [1-H(x)]K_{s_1}(x+X(t)-x_0)+H(x)[K_{s_2}(x-X(t)+x_{0})],
\end{equation}%
where $X(t)$ is the half-separation of the kink and antikink, $H(x)$ is the Heaviside function, and
\begin{equation}
K_{s_1}(x) = \left\{
        \begin{array}{ll}
            f_s(x), & \quad x \leq 0 \\
            f_s(0), & \quad x > 0
        \end{array}
    \right., \quad\text{and}\quad
  K_{s_2}(x) = \left\{
        \begin{array}{ll}
            f_s(0), & \quad x < 0 \\
            f_s(x), & \quad x \geq 0
        \end{array}
    \right..  
\end{equation}
Observe that when $X(0)=x_0$, we have $\varphi(x,0)=f_s(x)$, which implies that the initial conditions for the $\varphi^8$ field theory's equation of motion (``PDE model'') and the CC method (``ODE model'') both match. Using the ansatz in Eq.~\eqref{sp-cc},  and defining $K_{s_1}^+=K_{s_1}(x+X(t)-x_0)$ and $K_{s_2}^-=K_{s_2}(x-X(t)+x_0)$, we calculate the coefficient functions in Eq.~\eqref{lag1} as follows:
\begin{align}
b_0(X) &= \frac{1}{2}\int \left\{[1-H(x)]K_{s_1}'^+-H(x)K_{s_2}'^-\right\}^2 \, dx,\\
b_1(X) &= \frac{1}{2}\int \left\{[1-H(x)]K_{s_1}'^++H(x)K_{s_2}'^-\right\}^2  \, dx +\int V\Big([1-H(x)]K_{s_1}^++H(x)K_{s_2}^-\Big) \, dx.
\end{align}

\begin{figure}[tbp]
\begin{center}
\subfigure[]{
\includegraphics[scale=0.4]{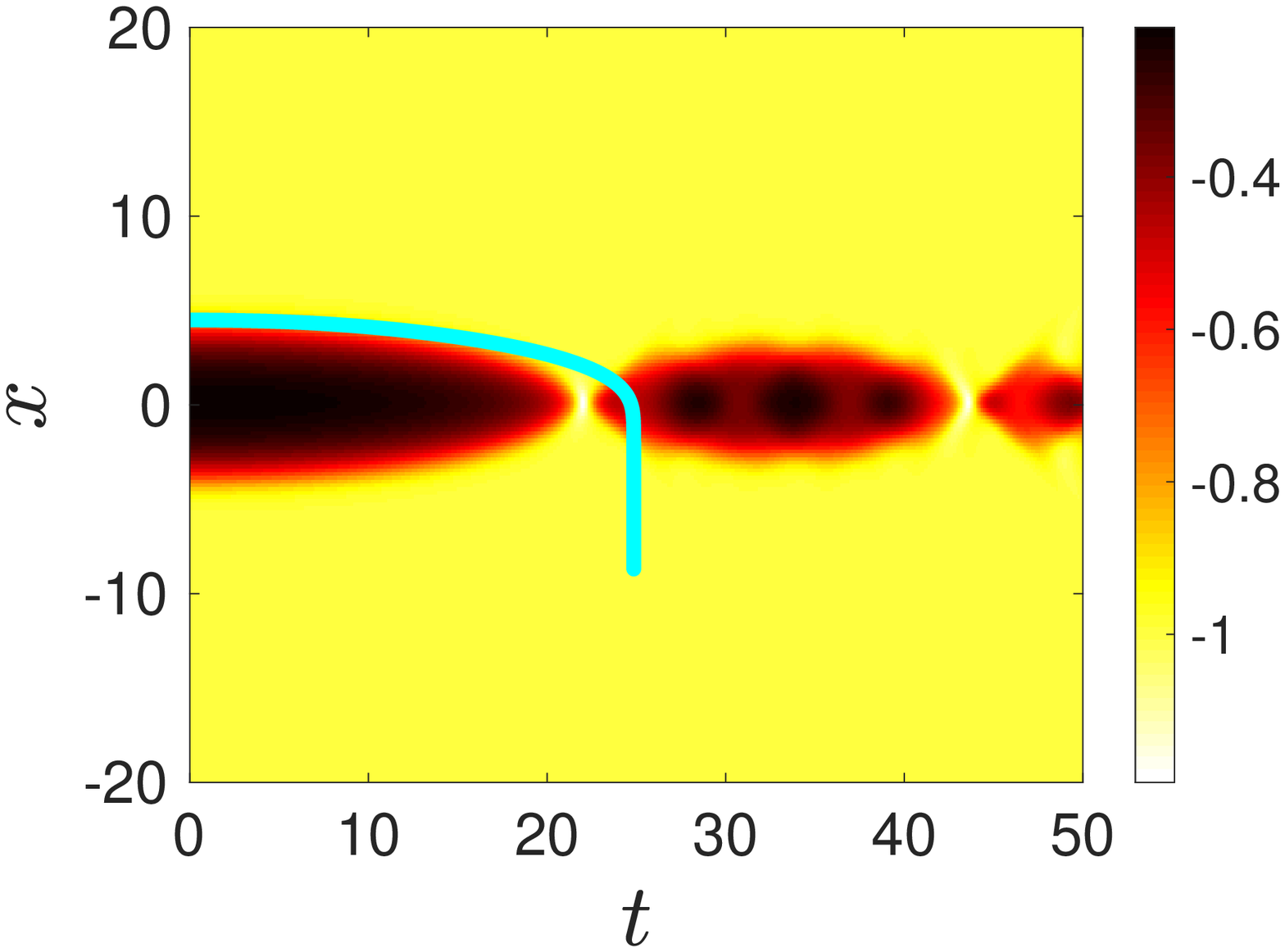}} 
\subfigure[]{
\includegraphics[scale=0.4]{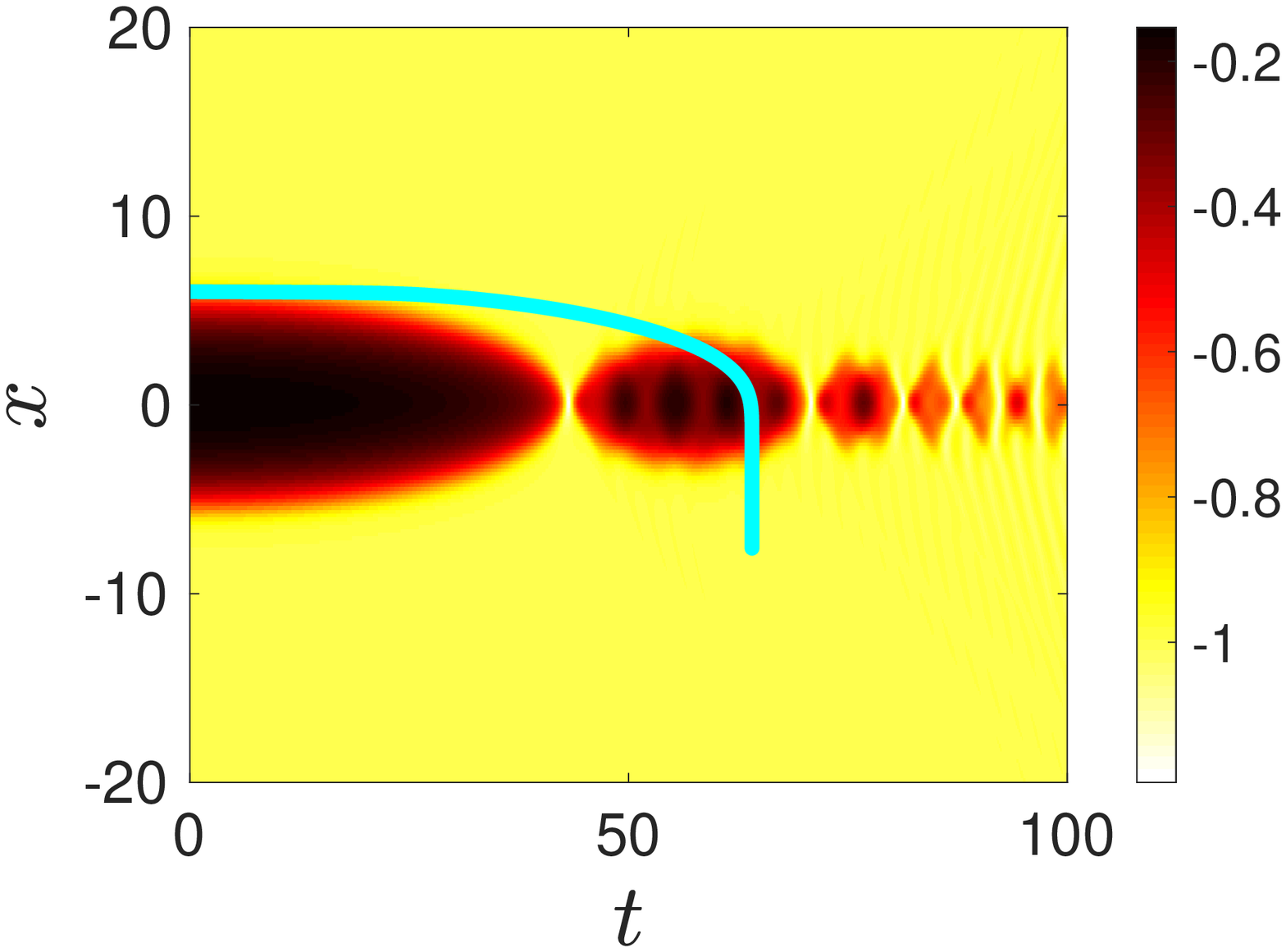}}
\end{center}
\caption{Using the minimized split-domain ansatz, overlays of the ODE model~\eqref{eq:EL} solution $X(t)$ (solid bold curve) on top of PDE model contour plot of $\varphi(x,t)$ for the evolution of an initially stationary ($v = 0$) kink-antikink configuration with (a) $x_0=4.5$ and (b) $x_0=6$.}
\label{sp-cc1}
\end{figure}

For zero initial velocity, both the ODE model and the PDE model show attraction when $x_0<x_c$,  where $6<x_c<7$. The ODE model's results agree better with the PDE results as $x_0$ becomes smaller. Figure~\ref{sp-cc1}(a) shows the ODE and PDE models' agreement, for $x_0=4.5$ and $v=0$, until the kink is expelled from the system in the ODE model. Meanwhile, Fig.~\ref{sp-cc1}(b) also shows attraction for $x_0=6$ and $v=0$; however, as before, the agreement between the ODE and PDE results is not as good.

\end{document}